\documentclass[preprint]{jpsj2}
\usepackage{graphicx}
\usepackage{amsmath}

\title{Models of universal power-law distributions}
\author{Kenji Kawamura* and Naomichi Hatano**}
\inst{*Department of Physics, Aoyama Gakuin University,
Chitosedai, Setagaya, Tokyo 157-8572

**Department of Physics, Aoyama Gakuin University,
Chitosedai, Setagaya, Tokyo 157-8572 

and

 Institute of Industrial Science, University of Tokyo, Komaba, Meguro, Tokyo 153-8505}
\abst{
Power-law distributions with various exponents are studied.
We first introduce a simple and generic model that reproduces Zipf's law.
We can regard this model both as the time evolution of the population of cities and that of the asset distribution.
We show that our model is very robust against various variations.
Next, we explain theoretically why our model reproduces Zipf's law.
By considering the time-evolution equation of our model, we see that the essence of Zipf's law is an asymmetric random walk in a logarithmic scale.
Finally, we extend our model by introducing an additional asymmetry.
We show that the extended model reproduces various power-law exponents.
By extending the theoretical argument for Zipf's law, we find a simple equation of the power-law exponent.
}
\recdate{}
\kword{Zipf's law, universality, power-law distribution}

\begin{document}
\maketitle

\section{Introduction}\label{intro}
Zipf's law is the observation that the frequency $n$ of the occurrence of various entities is an inverse power-law function $n\propto R^{-1}$, where the rank $R$ is determined by the descending order of the frequency of each entity.
Zipf \cite{Zipf} first made this observation for the frequency of occurrence of English words in literature;
that is, the most frequent word \lq\lq the'' (rank $R=1$) appears twice as many times as the second most frequent word \lq \lq of''(rank $R=2$).
Since then, Zipf's law has been found in various fields of natural and social sciences including the population of cities \cite{population} and the asset distribution of companies \cite{Takayasu,Takayasu2}.
This universality of Zipf's law, however, has not been well explained theoretically.
To date, most of the available models have been specific to each
problem\cite{population,Render,Sornette}.
One of the purposes of the present paper is to explain the mechanism of Zipf's law and thereby its universality.
Our conclusion is that the essence of Zipf's law is a size change of each entity proportional to its size.
(We made a brief report on this point previously \cite{kenji}.)

In various fields of natural and social sciences, there are also many phenomena that exhibit power laws with other exponents, $n\propto R^{-\kappa}$ with $\kappa\neq 1$, including internet \cite{internet}, traffic flow \cite{traffic}, economics, \cite{Pareto,Stanley,Stanley2} fish school \cite{fish} and family names \cite{family}.
We explain the mechanism of these power laws as well by extending the argument for Zipf's law.
The exponent differs from unity whenever the size increase and the size decrease are asymmetric.

In section \ref{model}, we introduce a simple and generic model that reproduces Zipf's law.
We can regard this model both as the time evolution of the population of cities and that of the asset distribution.
In section \ref{uni}, we explain theoretically why this model reproduces the power law. 
In section \ref{exmodel}, we extend our model so as to reproduce power laws with other power-law exponents.
Our explanation shows that the power-law of our model is very robust.

Before introducing our model, let us make a tutorial remark on the relation among three types of distribution appearing in the present paper:
the ranked distribution, the cumulative distribution function and the probability distribution function.
The ranked distribution, including Zipf's law, is a cumulative distribution function with the horizontal and vertical axes being flipped.
A ranked distribution
\begin{eqnarray}
n\propto R^{-\kappa}
\end{eqnarray}
means that there are, say, $R$ cities whose population is greater than or equal to $n$.
In other words, the probability that the population of an arbitrarily chosen city is greater than or equal to $n$ is given by
\begin{eqnarray}
P(\geq n)=\frac{R}{N},
\end{eqnarray}
where $N$ is the total number of cities.
Changing the variable $R$ to $n$ on the right-hand side, we have
\begin{eqnarray}
P(\geq n)\propto n^{-1/\kappa}=n^{-b},
\end{eqnarray}
where $b$ is the power-law exponent of the cumulative distribution function $P(\geq n)$.
Thus we obtain the relation between the power-law exponents of the cumulative distribution function and the ranked distribution.

Next, the cumulative distribution function $P(\geq x)$ is related to the probability distribution function $p(x)$ in the form 
\begin{eqnarray}\label{Pp}
P(\geq x)=\int_x^{\infty} p(x)dx.
\end{eqnarray}
This means that
\begin{eqnarray}\label{ab} 
p(x)\propto x^{-a} \quad \Leftrightarrow \quad  P(\geq x)\propto x^{-(a-1)}=x^{-b}.
\end{eqnarray}
where $a$ is the power-law exponent of the probability distribution function.

Consequently we arrive at the relation among the three power-law exponents as follows:
\begin{eqnarray}\label{exp} 
\frac{1}{\kappa}=b=a-1.
\end{eqnarray}
In the following sections, we introduce models that reproduce power laws in the form of the cumulative distribution function $P(\geq x)\propto x^{-b}$.
It corresponds to the power laws of the probability distribution function and the ranked distribution in the forms $p(x)\propto x^{-(b+1)}$ and $n\propto R^{-1/b}$, respectively.
In particular, the case $b=1$ is equivalent to Zipf's law in the form of the ranked distribution $n\propto R^{-1}$ .

In the present paper, we claim that the origin of the power laws is diffusion in the logarithmic scale.
For this reason, we also make a remark on the variable transformation of Eq~(\ref{ab}).
To see Eq.~(\ref{ab}) in the logarithmic scale, we change the variable to be $\xi=\log x$. 
The probability distribution function $p(x)$ is transformed to
\begin{eqnarray}
p(x)dx\sim x^{-a}dx=e^{-a\xi}e^\xi d\xi
               \equiv \tilde p(\xi)d\xi. 
\end{eqnarray}
In other words, we define the probability distribution function with respect to $\xi$ as
\begin{eqnarray}
\tilde p(\xi)=e^{-(a-1)\xi}=e^{-b\xi}.
\end{eqnarray}
It shows that the exponent of the exponential law of the probability distribution function in the logarithmic scale coincides with the power-law exponent of the cumulative distribution function.

\section{Model and simulation results}\label{model}
In this section we introduce our model and show the results of its simulation.
The model that we introduce here evolves as follows: 
First, we consider a set of positive values $\{x_i\}$ with $N$ entities.
Then we repeat the following procedures $T$ times:

\textit{Population-Distribution Model}
\begin{enumerate}
\renewcommand{\theenumi}{\roman{enumi}}
\item Choose an entity $i$ randomly from $1\leq i\leq N$. 
\item Add or subtract randomly the amount $\alpha x_i$ from the chosen entity with the probability $1/2$:
\begin{eqnarray}\label{xpm}
x_{i}\rightarrow x_{i}\pm\alpha x_{i},
\end{eqnarray}
\end{enumerate}
where $\alpha$ is a constant with $0<\alpha<1$.
For simplicity, we assume that the initial values of $\{x_{i}\}$ are all equal. 
This initial condition, however, is irrelevant in the long-time limit $T \rightarrow \infty$.

If we regard the values $\{x_i\}$ as the population of $N$ cities, an explanation of this model from the viewpoint of the population distribution is that bigger cities tend to have greater population movement.
During the time evolution we require the boundary condition that $x_i$ should not be less than a lower bound $x_\mathrm{l.b.}(>0)$;
When an operation such as $x_{i}\rightarrow x_i-\alpha x_i<x_\mathrm{l.b.}$ is chosen in the step (ii), we cancel this operation.
This boundary condition is required in order to prevent each entity $x_{i}$ from decreasing limitlessly.
This is quite a plausible requirement;
in realities, the population of cities, for example, cannot be less than $x_\mathrm{l.b.}=1$.

\begin{figure}
\begin{center}
\includegraphics[width=0.7\textwidth]{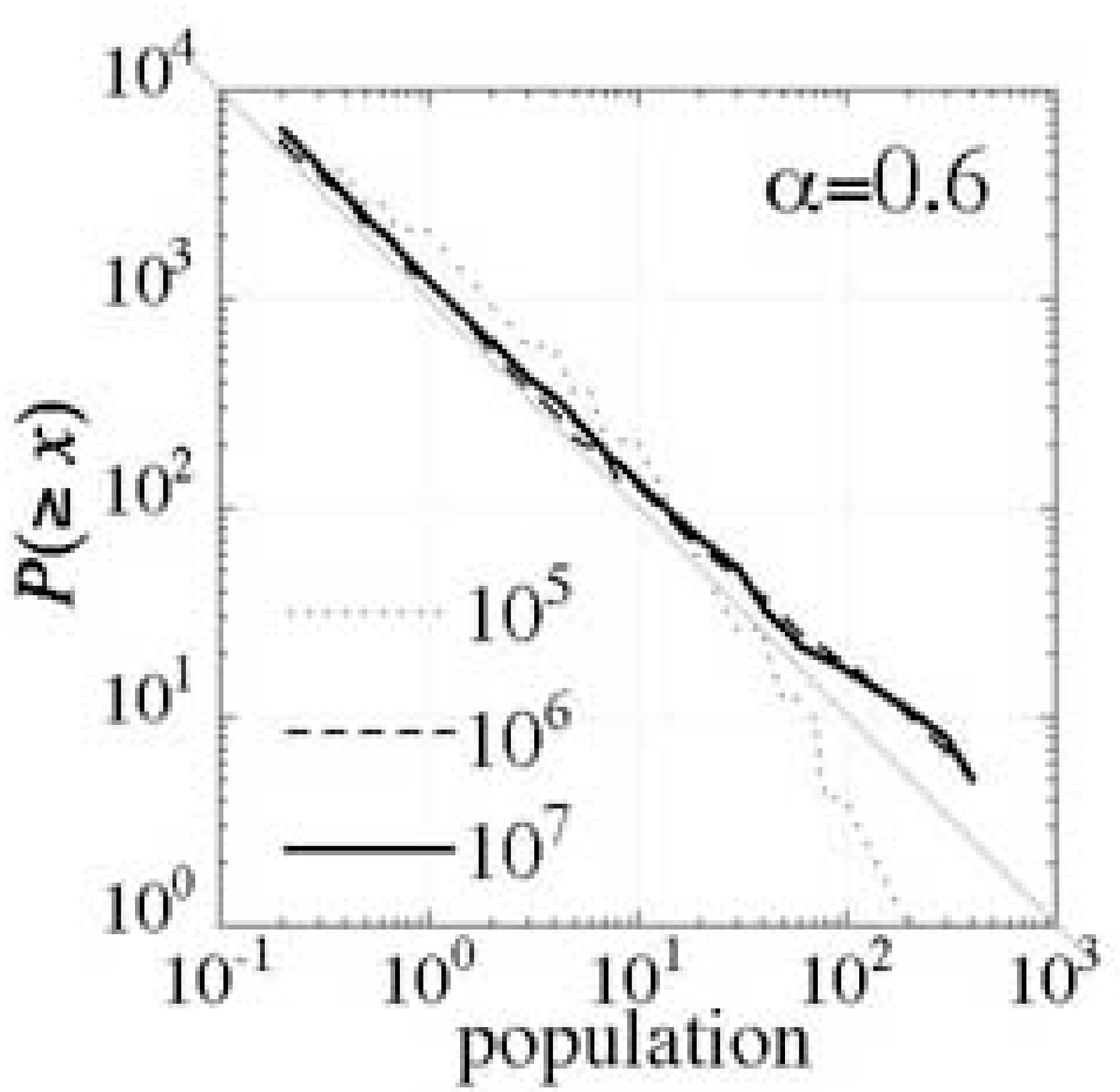}
\includegraphics[width=0.7\textwidth]{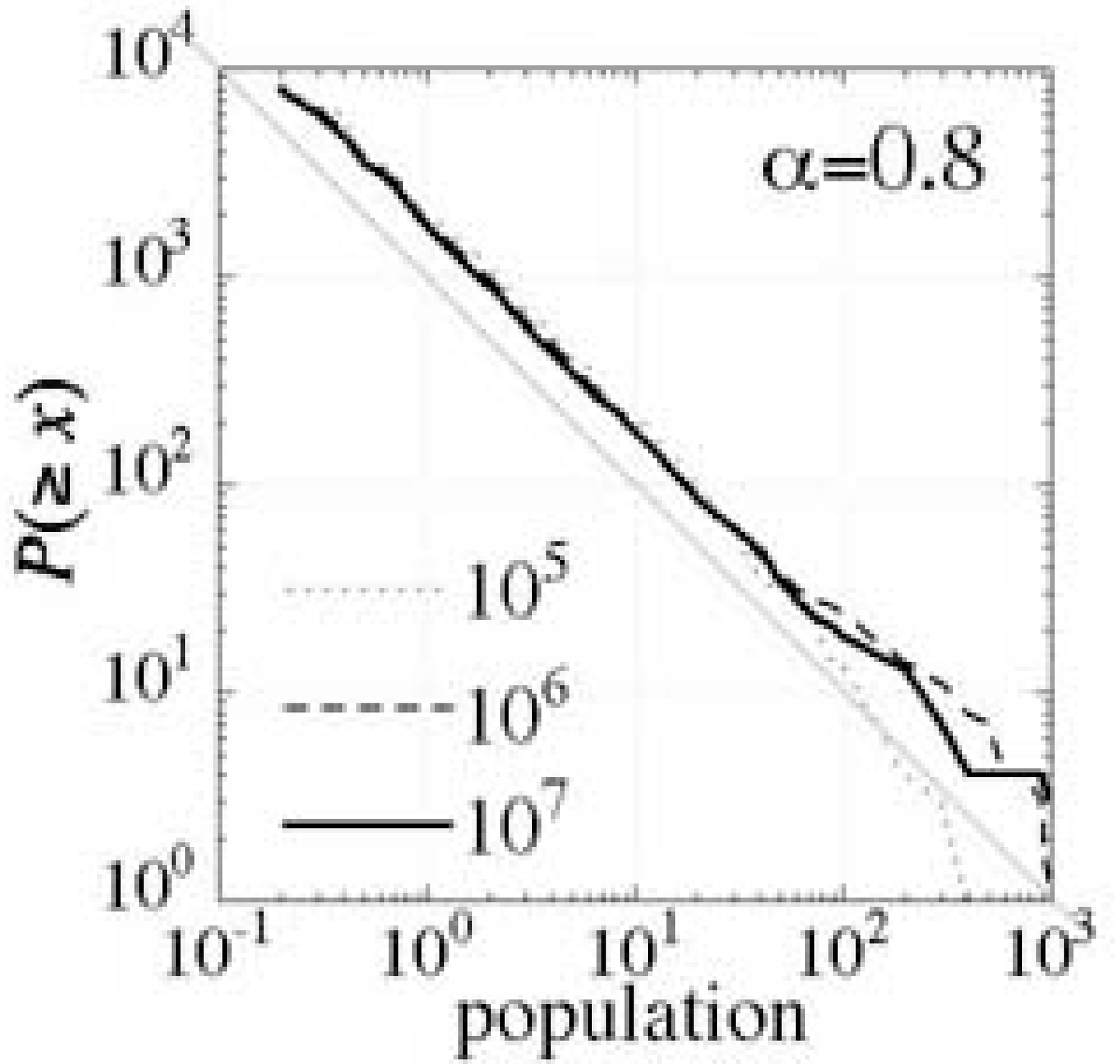}
\caption{Results of the simulation for various values of $\alpha$.
 The number of steps is $T=$ $10^5$ (dotted line), $10^6$ (dashed line) and $10^7$ (solid line). 
The straight line indicates the power law with the exponent $b=1$, namely Zipf's law.
}\label{zipf}
\end{center}
\end{figure}

\begin{figure}
\begin{center}
\includegraphics[width=0.7\textwidth]{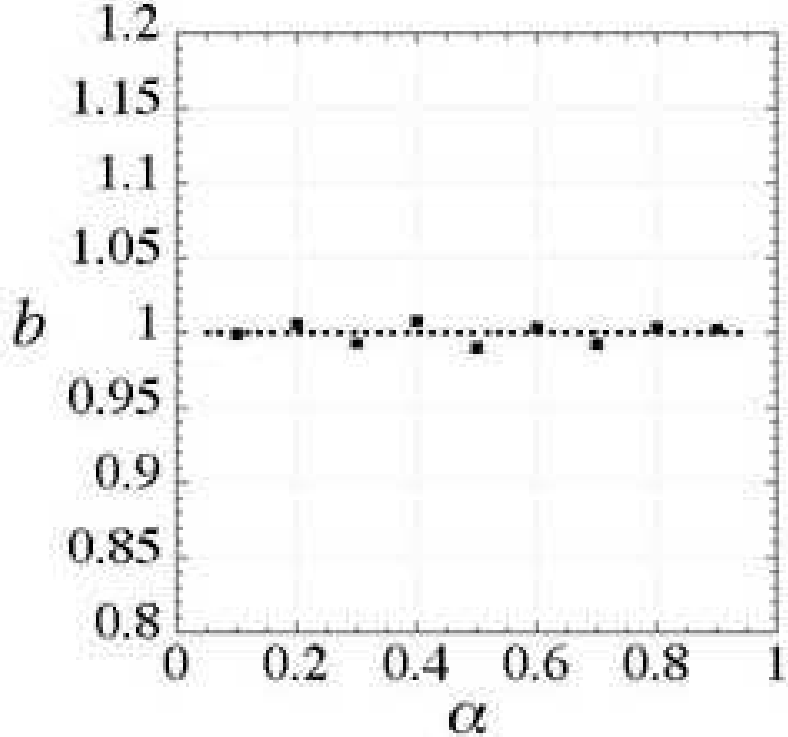}
\caption{The power-law exponent $b$ against the parameter $\alpha$.
The straight line shows the solution Eq.~(\ref{fb}), while the squares indicate estimates from the results of our simulation.}
\label{alpha}
\end{center}
\end{figure}

Figure \ref{zipf} shows some of the results of the simulation of the above procedures for various values of $\alpha$ in the form of the cumulative distribution function $P(\geq x)\propto x^{-b}$. 
(Here the number of the entities is $N=10^4$.)
\begin{figure}
\begin{center}
\includegraphics[width=0.7\textwidth]{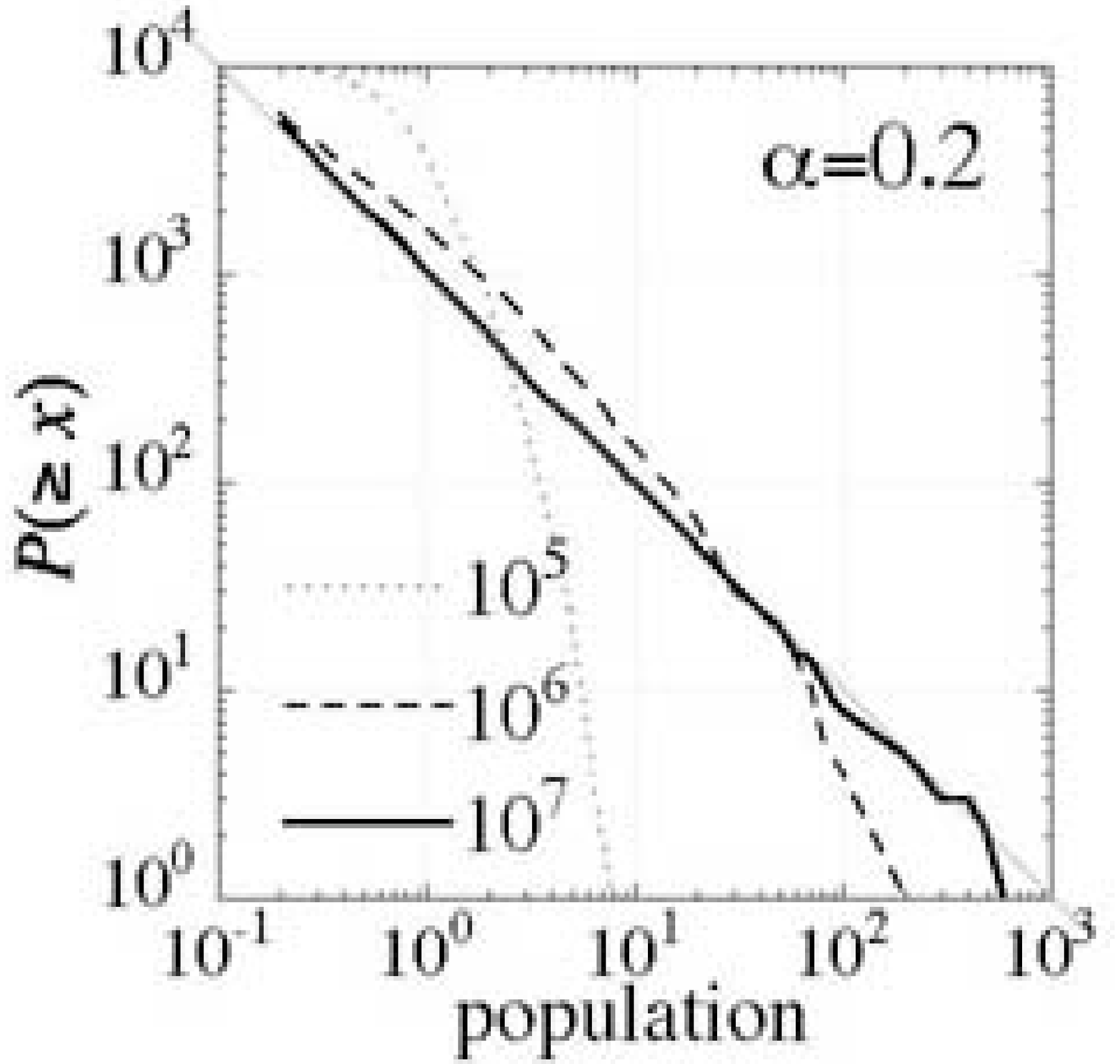}
\includegraphics[width=0.7\textwidth]{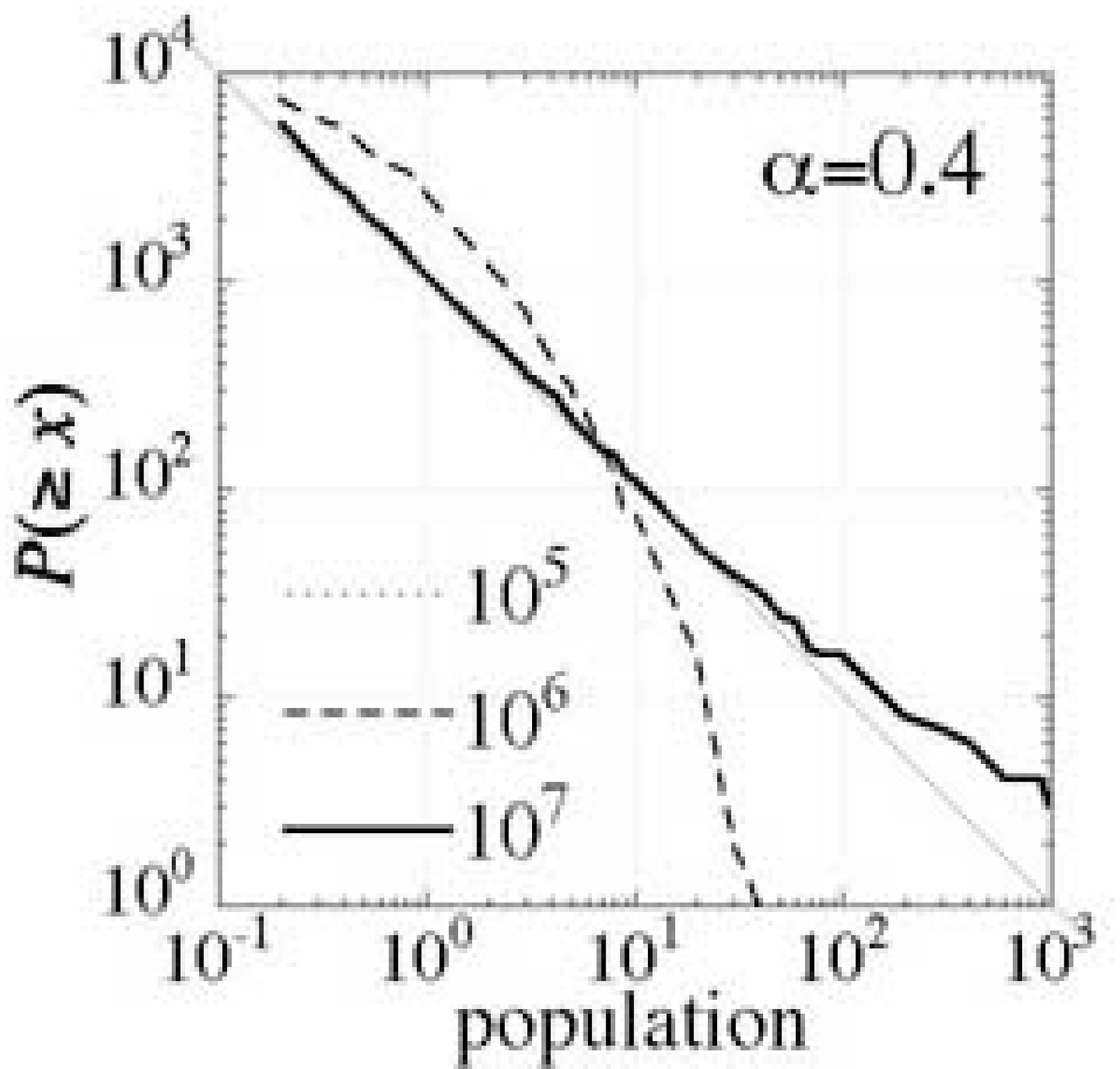}
\end{center}
\end{figure}%
We can clearly see that the distribution converges to power laws in all cases in the long-time limit $T\rightarrow \infty$.
Figure~\ref{alpha} shows the power-law exponents estimated from the results of our simulation.
(We estimated the exponents by the least-squares method.)
The exponent $b$ is obviously close to unity for any values of $\alpha$.
We note that in Fig.~\ref{zipf} the parameter $\alpha$ is related only to the rapidity of the convergence to Zipf's law.
Because of the relation (\ref{exp}), we obtain Zipf's law in the form of the ranked distribution $n\propto R^{-1}$ by flipping the horizontal and vertical axes of  Fig.~\ref{zipf};
the model thus reproduces the universality of Zipf's law as $T \rightarrow \infty$.

\begin{figure}
\begin{center}
\includegraphics[width=0.7\textwidth]{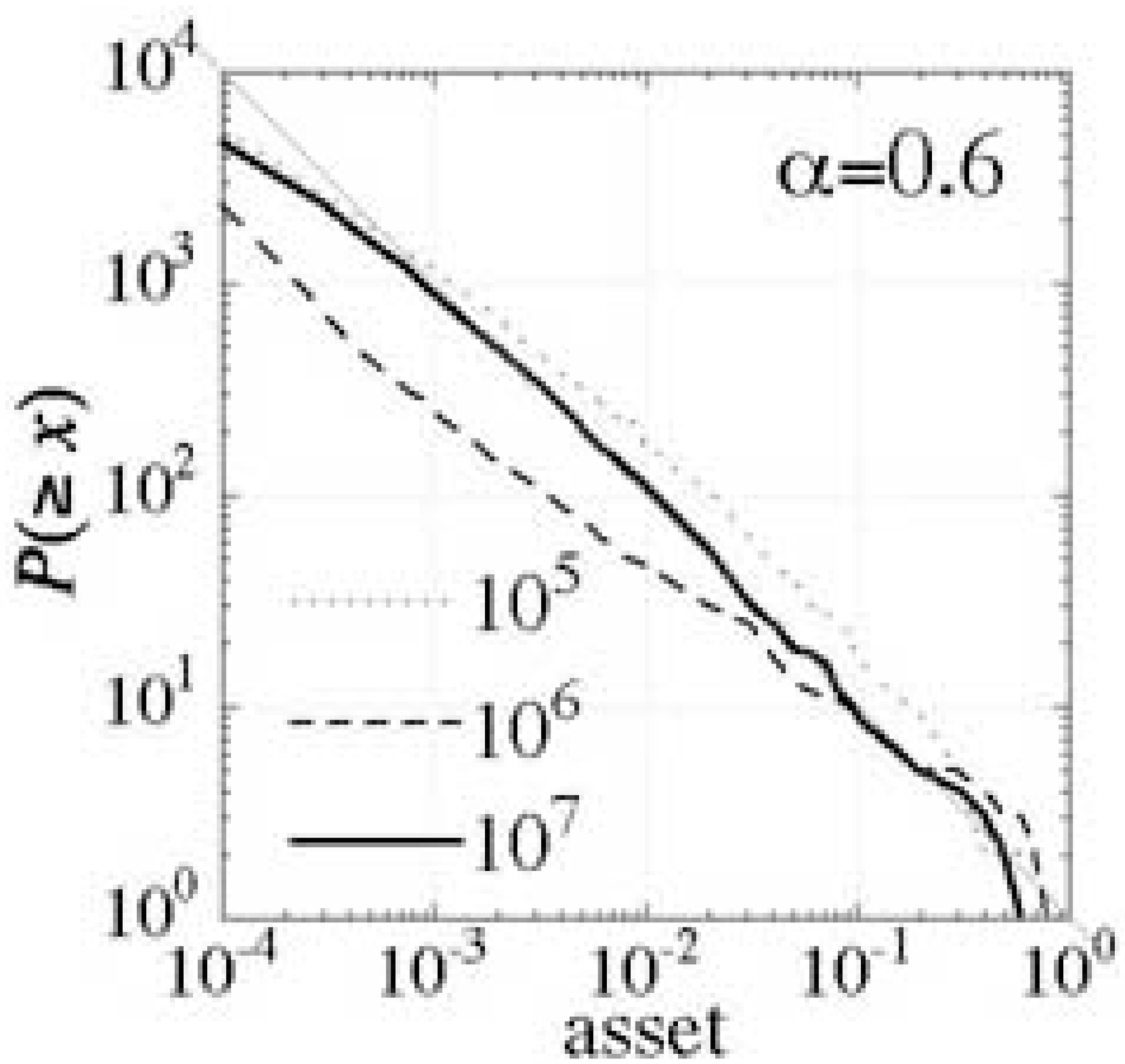}
\includegraphics[width=0.7\textwidth]{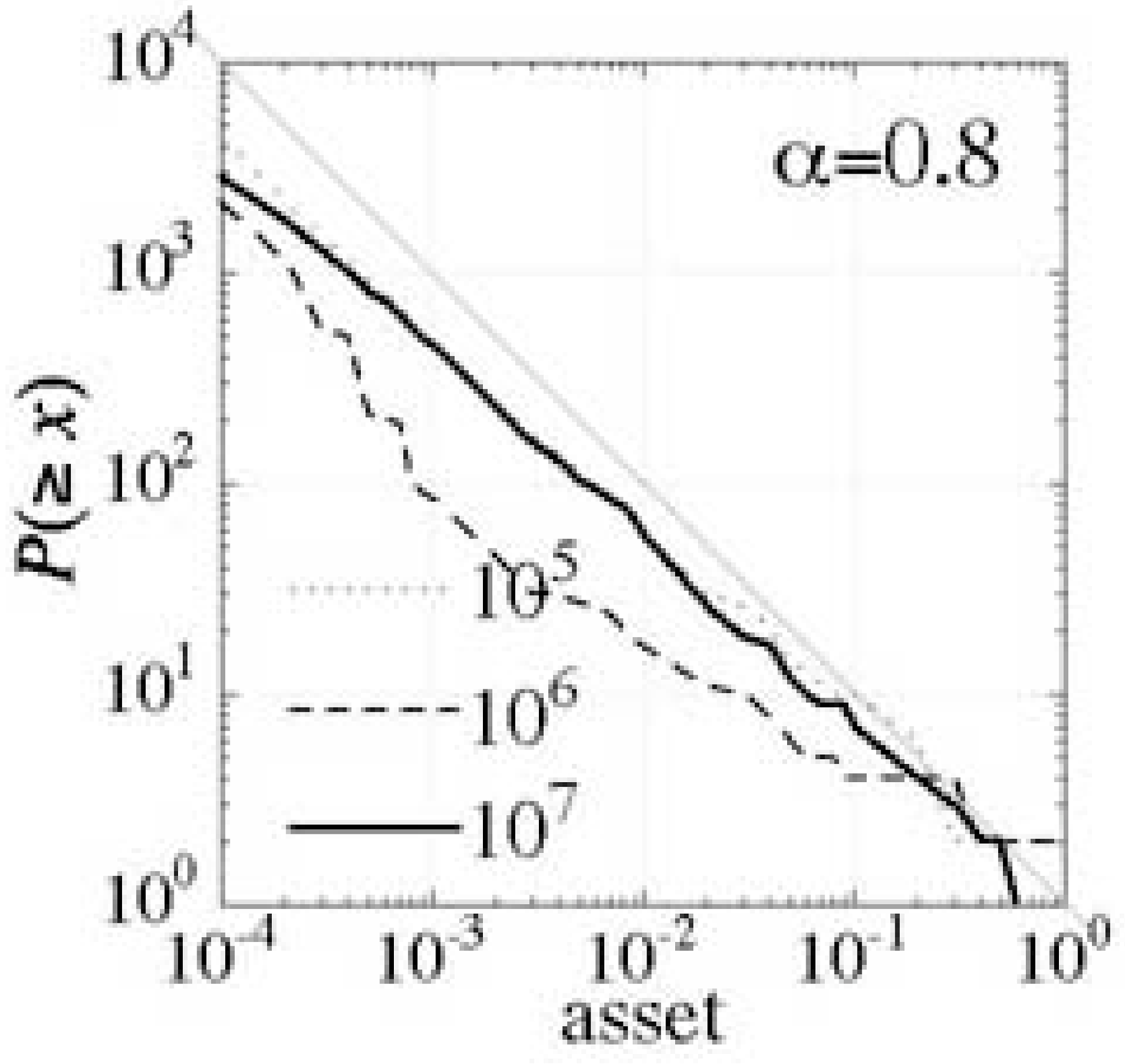}
\caption{Results of the simulation of the asset distribution for various values of $\alpha$.
 The number of steps is $T=$ $10^5$ (dotted line), $10^6$ (dashed line) and $10^7$ (solid line). 
 We normalized the results by the asset of the highest entity.
The straight line indicates the power law with the exponent $b=1$, namely Zipf's law.}
\label{company-zipf}
\end{center}
\end{figure}

To show the robustness of the model, we modify it from the viewpoint of the asset distribution and still reproduce Zipf's law.
Assume that a company tends to trade with a company of a similar size.
Then we repeat the following procedures $T$ times:

\textit{Asset-Distribution Model}
\begin{enumerate}
\renewcommand{\theenumi}{\roman{enumi}}
\item Choose an entity $i$ randomly from $1\leq i\leq N$. 
\item For $i>1$, we move the amount $\alpha x_{i-1}$ from the $(i-1)$th entity to the $i$th entity, where $\alpha$ is a constant parameter with $0<\alpha <1$.
In other words, the following takes place:
\begin{equation}
x_{i-1}\rightarrow x_{i-1}-\alpha x_{i-1} \label{xm}, 
\end{equation}
\begin{equation}
x_{i}\rightarrow x_{i}+\alpha x_{i-1}. \label{xp}
\end{equation}
For $i=1$, we increase all values $x_{i}$ by the same amount $x_{1}\alpha/N$.
\item Rearrange the entities in the descending order of the values $x_{i}$. 
\end{enumerate}
An intuitive explanation of this modified model is that money flows from one company to a smaller company and the smaller company delivers products to the larger company.
Figure \ref{company-zipf} shows some of the results of the simulation of the above procedures for various values of $\alpha$.
(Here the number of the companies is $N=10^4$.)
\begin{figure}
\begin{center}
\includegraphics[width=0.7\textwidth]{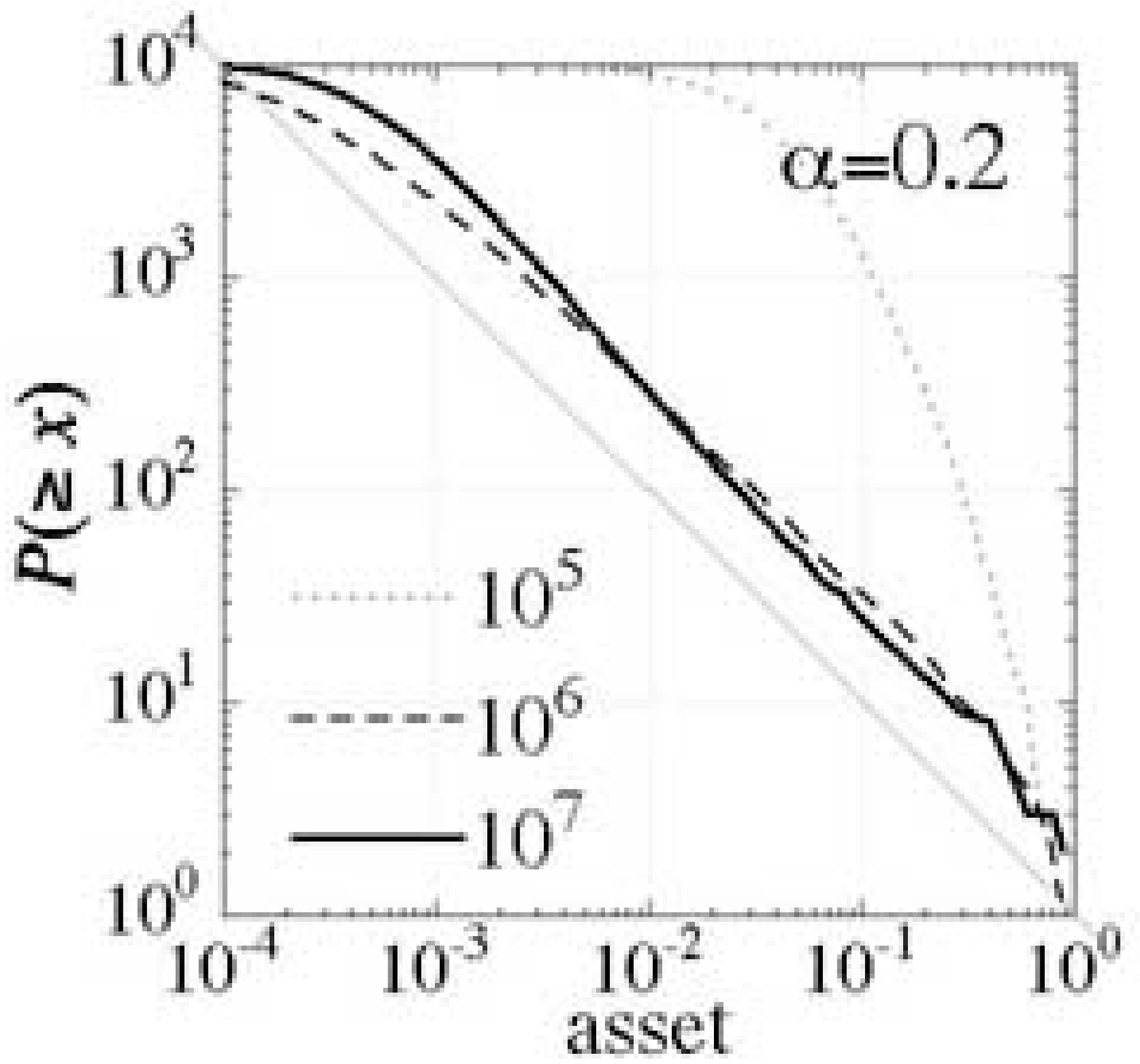}
\includegraphics[width=0.7\textwidth]{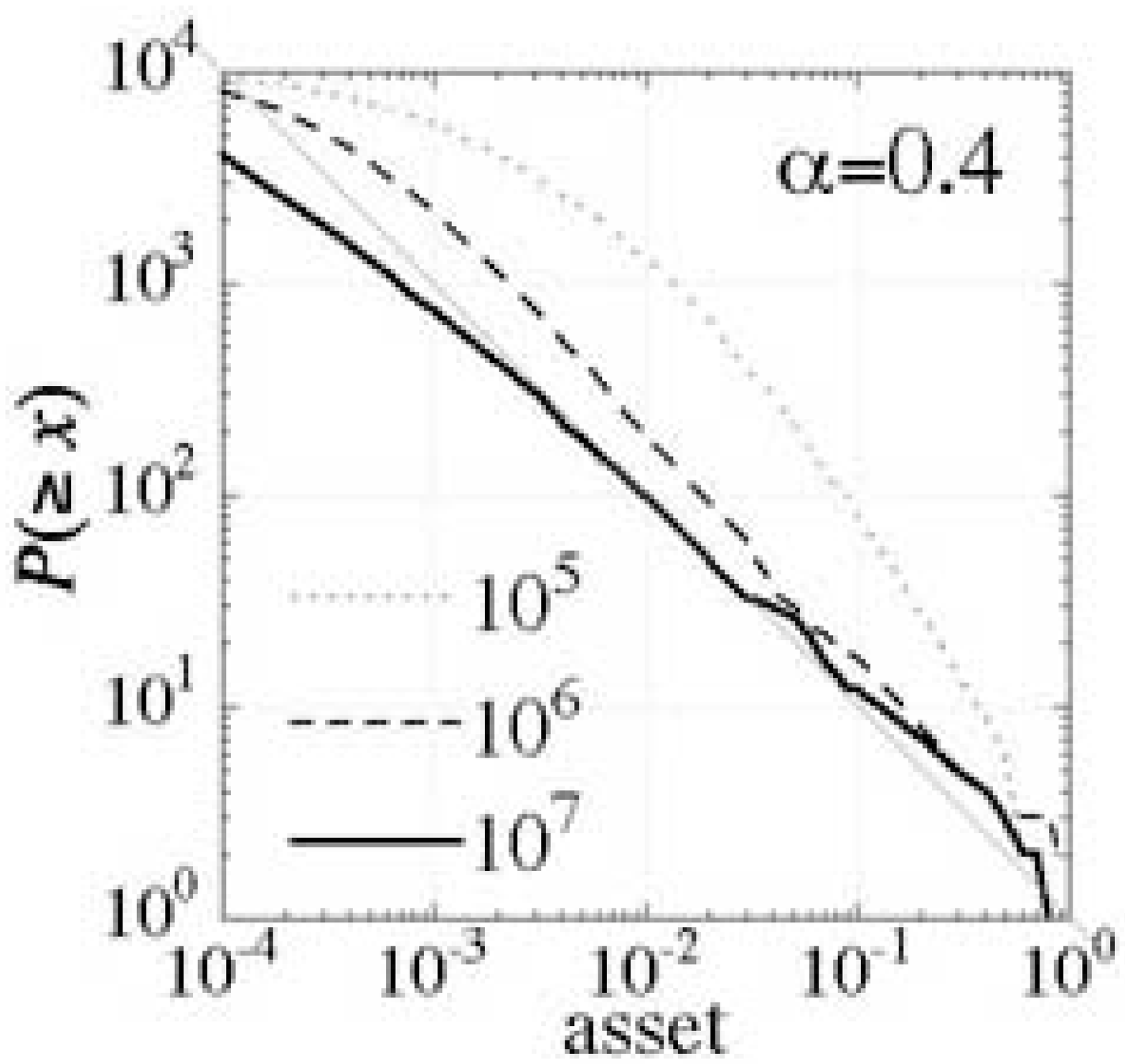}
\end{center}
\end{figure}%
The cumulative distribution function again converges to $P(\geq x)\propto x^{-1}$ in all cases.
The operation for $i=1$ in the step (ii) corresponds to the boundary condition $x_i \geq x_{\mathrm{l.b.}}$ of the population-distribution model.

\begin{figure}
\begin{center}
\includegraphics[width=0.7\textwidth]{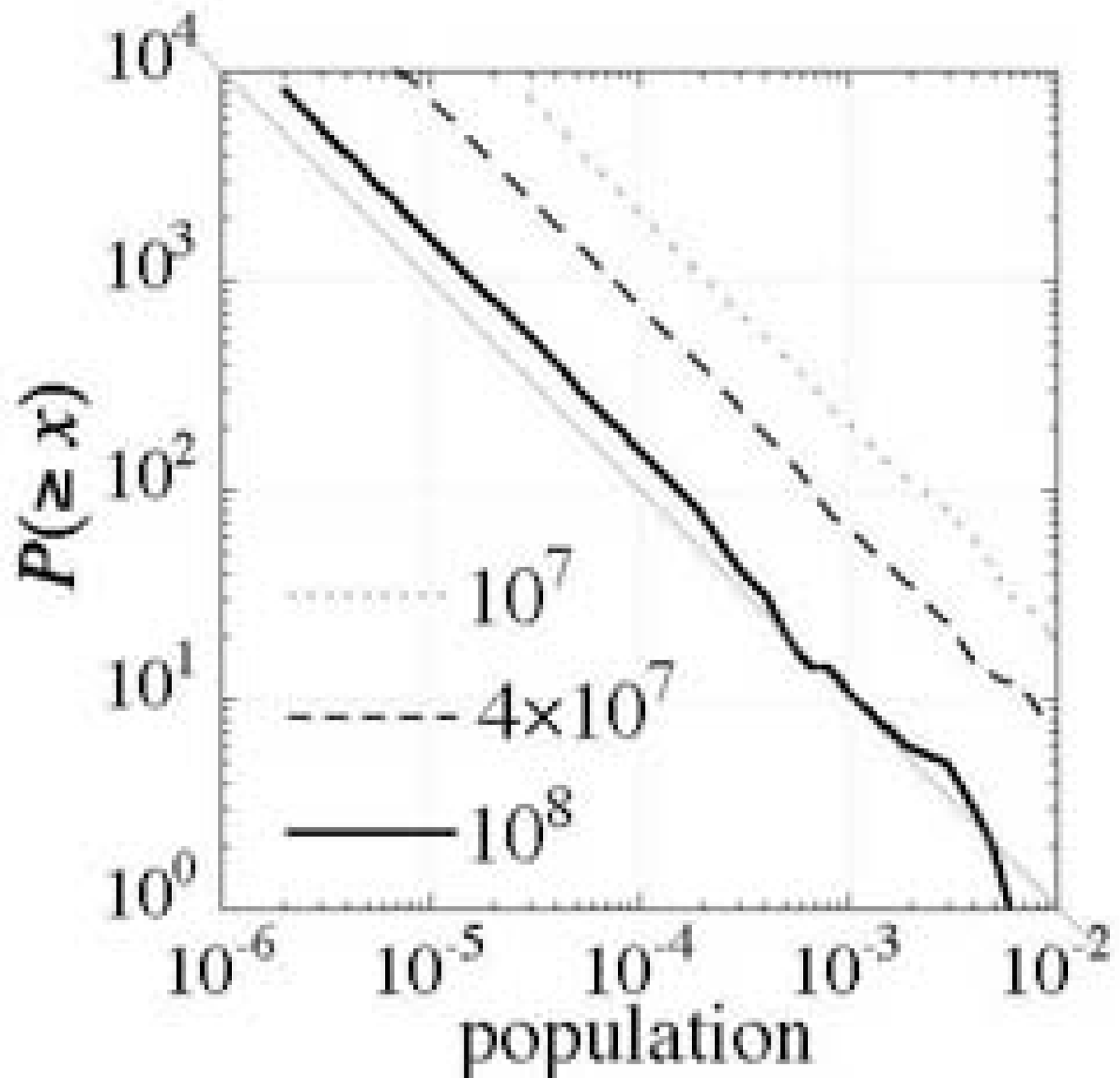}
\caption{The result of the simulation for $\alpha=0.3$ with a broad lower bound.
 The number of steps is $T=$ $10^7$ (dotted line), $4\times10^7$ (dashed line) and $10^8$ (solid line). 
The straight line indicates the power law with the exponent $b=1$, namely Zipf's law.}
\label{boundary}
\end{center}
\end{figure}

The value of the lower bound $x_\mathrm{l.b.}$ does not need to be fixed.
Let us modify the population-distribution model and make $x_\mathrm{l.b.}$ a changeable parameter.
When the operation $x_{i}\rightarrow x_i-\alpha x_i<x_\mathrm{l.b.}$ is chosen in the step (ii) of the first model, we produce a random number $S$ ($0<S<1$).
For 
\begin{equation}
e^{-(x_\mathrm{l.b.}-x_i)}>S,
\end{equation}
this operation is cancelled, but for
\begin{equation}
e^{-(x_\mathrm{l.b.}-x_i)}<S,
\end{equation}
$x_i$ becomes the new value of $x_\mathrm{l.b.}$.
A broad change that exceeds the lower bound $x_\mathrm{l.b.}$ is allowed with a small probability.
Figure~\ref{boundary} shows the result of the simulation of the modified model for $\alpha=0.3$ and $N=10^4$.
Compare it with Fig.~\ref{zipf}; we can see the distribution in Fig.~\ref{boundary} behaves in the form of the Zipf's law, while shifting to the negative direction.

\begin{figure}
\begin{center}
\includegraphics[width=0.7\textwidth]{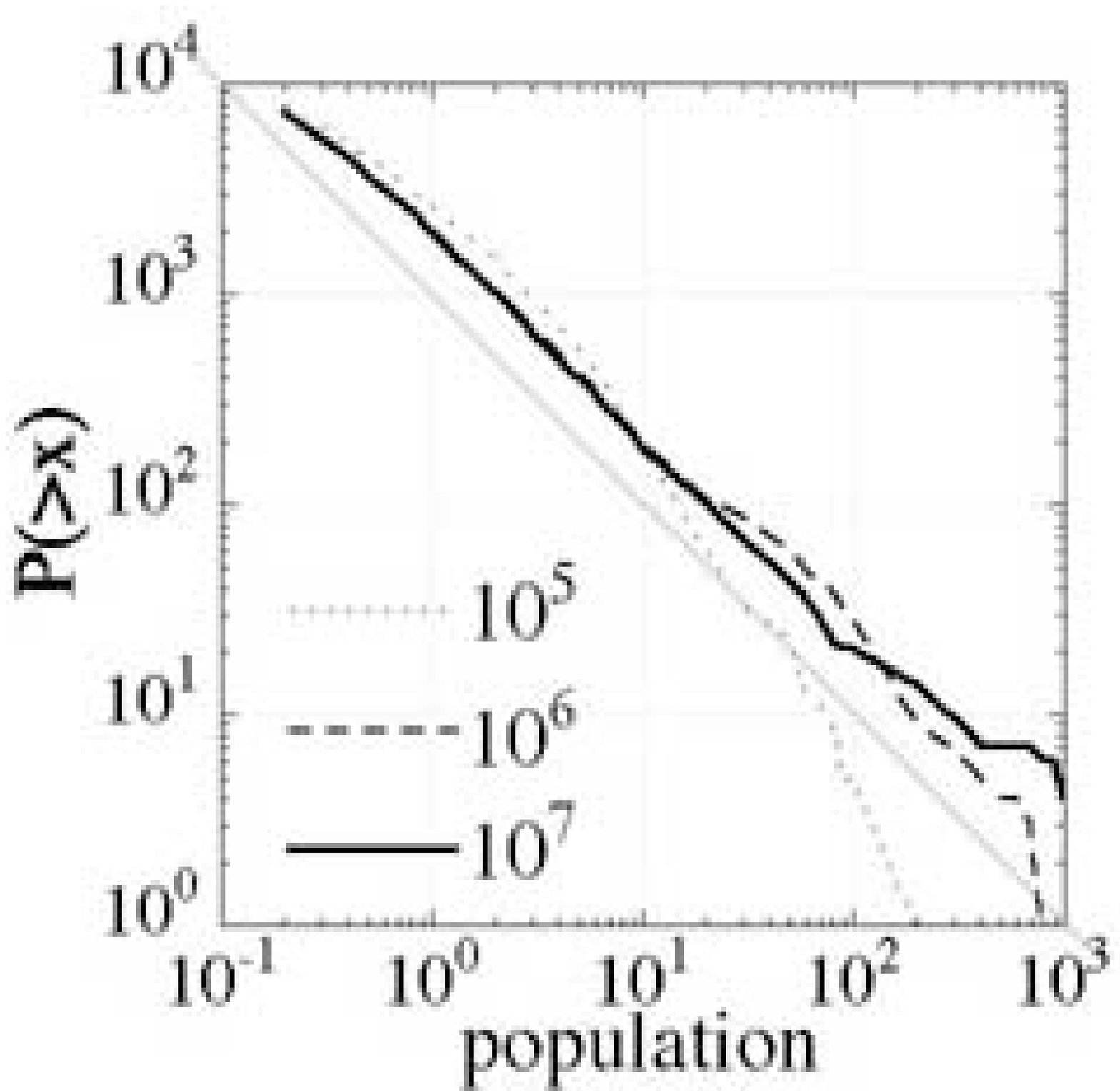}
\caption{The result of the simulation with randomly chosen $\alpha$ ($0<\alpha<1$).
 The number of steps is $T=$ $10^5$ (dotted line), $10^6$ (dashed line) and $10^7$ (solid line). 
The straight line indicates the power law with the exponent $b=1$, namely Zipf's law.}
\label{ransu}
\end{center}
\end{figure}

The value $\alpha$ does not need to be fixed as well.
We can modify the population-distribution model and make $\alpha$ as a changeable parameter;
we choose $\alpha$ randomly from $0<\alpha<1$ every time between the steps (i) and (ii) in the population-distribution model.
In other words, the value of $\alpha$ changes at each time step.
Figure~\ref{ransu} shows the result of the simulation of the latter modified model for $N=10^4$.
Compare it with Fig.~\ref{zipf}; we can see the distribution in Fig.~\ref{ransu} behaves in the form of the Zipf's law.
The results in Figs.~\ref{boundary} and \ref{ransu} show that Zipf's law of our model holds out against a wide variety of modification.

We see from the above variations that the essence of the model is the increase and the decrease proportional to the size of each entity;
this is the key to the universality of Zipf's law.
On the basis of the simpleness and the robustness of our model, we believe that Zipf's law in many fields of science has the same origin of \lq\lq proportional change."

\section{Continuum Limit and Universality}\label{uni}
In this section we give a theoretical explanation as to why the model in the previous section reproduces Zipf's law.
Let us consider the limit $N \rightarrow \infty$.
Hence we regard the set of variables $\{x_{i}\}$ as a continuous variable $x$.
We consider in this limit the time evolution of the probability distribution function $p(x)$.
We note that Eq.~(\ref{xpm})  gives probability flows in the directions
\begin{eqnarray}\label{evolve}
x \rightarrow x\pm \alpha x=(1\pm \alpha)x.
\end{eqnarray}
The time-evolution equation of the probability distribution function $p(x)$ is therefore given by the following:
\begin{equation}\label{x}
\frac{\partial}{\partial t}p(x,t)=-\gamma p(x,t)+\frac{1}{2}\frac{\gamma}{1-\alpha}p\left(\frac{x}{1-\alpha},t\right)+\frac{1}{2}\frac{\gamma}{1+\alpha}p\left(\frac{x}{1+\alpha},t\right),
\end{equation}
where $p(x,t)$ is the probability distribution of $x$ at time $t$ and $\gamma $ is a  constant determining the unit of time. 
The first term on the right-hand side represents a flow from the point $x$ out to the points $(1\pm \alpha)x$, while the second and third terms represents flows from $x/(1 \pm \alpha)$ into $x$.

The coefficients $1/(1 \pm \alpha)$ in front of $p$ in Eq.~(\ref{x}) are necessary in order to satisfy the probability conservation.
To see this, we integrate Eq.~(\ref{x}) over $x$ as
\begin{eqnarray}\label{int}
\frac{\partial}{\partial t} \int_0^\infty p(x,t)dx&=&-\gamma \int_0^\infty p(x,t)dx 
+\frac{1}{2}\frac{\gamma}{1-\alpha}\int_0^\infty p\left(\frac{x}{1-\alpha},t\right)dx \nonumber \\ 
&+&\frac{1}{2}\frac{\gamma}{1+\alpha}\int_0^\infty p\left(\frac{x}{1+\alpha},t\right)dx.
\end{eqnarray}
Changing the variables to be
\begin{equation}
x' =\frac{x}{1-\alpha}, \qquad
x''=\frac{x}{1+\alpha},
\end{equation}
we can rewrite Eq.~(\ref{int}) as
\begin{eqnarray}
\frac{\partial}{\partial t} \int_0^\infty p(x,t)dx&=&-\gamma \int_0^\infty p(x,t)dx
+\frac{\gamma}{2} \int_0^\infty p(x',t)dx' \nonumber \\
&+&\frac{\gamma}{2} \int_0^\infty p(x'',t)dx'' \nonumber 
= 0.
\end{eqnarray}
Thus the total probability is conserved.
(In the present case, the probability would be conserved without the coefficients $1/(1\pm \alpha)$.
However, they are necessary when Eq.~(\ref{x}) is generalized later; see Eq.~(\ref{r}).)

Let us see Eq.~(\ref{x}) in the logarithmic scale; we change the variable to be $\xi=\log x$. 
The probability distribution function is transformed to
\begin{eqnarray}
p(x,t)dx=p(e^\xi,t)e^\xi d\xi= \tilde p(\xi,t)d\xi.
\end{eqnarray}
In other words, we define
\begin{eqnarray}
\tilde p(\xi,t)\equiv p(x,t)x=p(e^\xi,t)e^\xi.
\end{eqnarray}
By rewriting $p$ in terms of $\tilde p$ as
\begin{eqnarray}
p(x,t)&=&\frac{1}{x} \tilde p(\log x,t)
=e^{-\xi} \tilde p(\xi,t),\\
p\left(\frac{x}{1 \mp \alpha},t\right)&=&\frac{1 \mp \alpha}{x}  \tilde p\left(\log \frac{x}{1 \mp \alpha},t\right)\nonumber\\
&=&(1 \mp \alpha)e^{-\xi}\tilde p(\xi -\log (1 \mp \alpha),t),
\end{eqnarray}
we transform the evolution equation (\ref{x}) to 
\begin{eqnarray}\label{xi}
\frac{\partial}{\partial t}\tilde{p}(\xi,t)&=&-\gamma\tilde p(\xi,t)+\frac{\gamma}{2}\tilde p(\xi+\beta_{-},t)+\frac{\gamma}{2}\tilde p(\xi-\beta_{+},t),  
\end{eqnarray}
where
\begin{eqnarray}\label{beta}
\beta_{\pm} &\equiv& \pm \log(1\pm \alpha)>0.
\end{eqnarray}
We can regard the evolution equation (\ref{xi}) as a random walk from $\xi$ to $\xi \pm \beta_{\pm}$, that is, a random walk with asymmetrically fixed step sizes. 

Let us consider the meaning of the asymmetric random walk.
We carry out the Taylor expansion of the second and third terms on the right-hand side of Eq. (\ref{xi}). 
Assuming $O(\beta_{\pm}^3)\simeq0$, we can rewrite Eq.~(\ref{xi}) as
\begin{equation}\label{Taylor}
\begin{array}{ll}
\frac{\partial}{\partial t}\tilde{p}(\xi,t)
=-2\gamma \tilde p(\xi,t)
&+\gamma \left(\tilde p(\xi,t)+\beta_{-}\frac{\partial}{\partial \xi}\tilde p(\xi,t)+\frac{\beta_{-}^2}{2} \frac{\partial^2}{\partial \xi^2}\tilde p(\xi,t) \right) \\
&+\gamma \left(\tilde p(\xi,t)-\beta_{+}\frac{\partial}{\partial \xi}\tilde p(\xi,t) +\frac{\beta_{+}^2}{2}\frac{\partial^2}{\partial \xi^2}\tilde p(\xi,t)  \right). 
\end{array}
\end{equation}
By rearranging the right-hand side of Eq.~(\ref{Taylor}), we have
\begin{equation}\label{advdif}
\frac{\partial}{\partial t}\tilde{p}(\xi,t)
=\gamma \frac{\beta_{+}^2+\beta_{-}^2}{2}\frac{\partial^2}{\partial \xi^2}\tilde p(\xi,t)
+\gamma (-\beta_{+}+\beta_{-})\frac{\partial}{\partial \xi}\tilde p(\xi,t).
\end{equation}
The first and second terms on the right-hand side of Eq.~(\ref{advdif}) correspond to the diffusion and advection terms, respectively.
Consequently we can regard the equation of the asymmetric random walk, Eq.~(\ref{xi}), as an advection-diffusion equation (\ref{advdif}) approximately.
For $0<\alpha <1$, the advection coefficient is
\begin{equation}
-\beta_{+}+\beta_{-}=-\log(1+\alpha) -\log (1-\alpha)=\log \frac{1}{1-\alpha^2}>0.
\end{equation}
It suggests that the probability distribution function in the logarithmic scale $\tilde p(\xi,t)$ moves to the negative direction while diffusing.

\begin{figure}
\begin{center}
\includegraphics[width=0.7\textwidth]{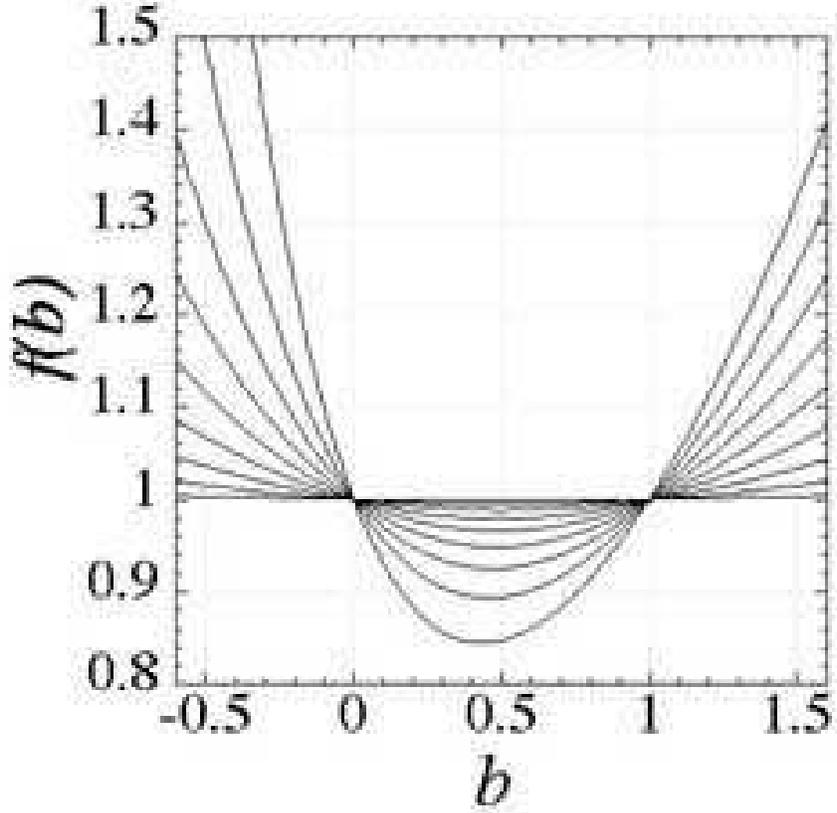}
\caption{
The function $f(b)$ defined in Eq.~(\protect\ref{fb}) for various values of $\alpha$.
The curvature of $f(b)$ increases as $\alpha$ becomes greater.
}
\label{fbalpha}
\end{center}
\end{figure}

Let us now concentrate on finding stationary solutions $\tilde p_s(\xi)$ of Eq.~(\ref{xi}):
\begin{eqnarray}\label{ss}
-2\gamma \tilde p_{s }(\xi)+\gamma \tilde p_{s }(\xi+\beta_{-})+\gamma \tilde p_{s }(\xi-\beta_{+})=0.
\end{eqnarray}
We seek solutions of the form
\begin{eqnarray}\label{sp}
\tilde p_{s}\propto e^{-b\xi}.
\end{eqnarray}
(See Appendix A for the validity of this assumption.)
By plugging Eq.~(\ref{sp}) into Eq.~(\ref{ss}), we have
\begin{eqnarray}\label{steady}
-2\gamma e^{-b\xi}+\gamma e^{-b\xi}e^{-b\beta_{-}}+\gamma e^{-b\xi}e^{b\beta_{+}}=\gamma e^{-b\xi}(-2+e^{-b\beta_{-}}+e^{b\beta_{+}} )=0,
\end{eqnarray}
or
\begin{eqnarray}\label{eb}
e^{b\beta_{+}}+e^{-b\beta_{-}}=2.
\end{eqnarray}
Rewriting Eq.~(\ref{eb}) using Eq.~(\ref{beta}), we obtain the following equation for $b$:
\begin{eqnarray}\label{fb}
f(b)\equiv\frac{1}{2}(1+\alpha)^b+\frac{1}{2}(1-\alpha)^b=1.
\end{eqnarray}
Figure~\ref{fbalpha} shows the function $f(b)$ for various values of $\alpha$.
Though the form of $f(b)$ changes with $\alpha$, the two solutions of Eq.~(\ref{fb}),  $b=0$ and $1$, are stable.
Obviously $b=0$ is a solution of Eq.~(\ref{fb}).
The solution $b=0$ means that the stationary solution $\tilde p_{s}$ is a uniform distribution:
\begin{equation}\label{const}
\tilde p_{s}\propto e^{-b\xi}=\mathrm{const}.
\end{equation}
As we suggested in Eq.~(\ref{advdif}), our model can be regarded as diffusion with advection in the logarithmic scale.
We hence expect that the solution $b=0$, or a uniform distribution appears only in the case of periodic boundary conditions.
For our model with the lower bound $x_\mathrm{l.b.}$, the solution $b=1$ survives:
\begin{equation}\label{ps1}
\tilde p_s \propto e^{-\xi}
\end{equation} 

According to the argument at the end of Sec.~\ref{intro}, the solution (\ref{ps1})  means $P(\geq x) \propto x^{-1}$.
This is in good agreement with our simulation data in Fig.~\ref{zipf}.
The present argument indicates that a random walk with asymmetric step sizes in the logarithmic scale is the essence of the universality of Zipf's law.

\section{Extension of the model}\label{exmodel}
So far, we have focused on Zipf's law ($b=1$). 
However, various power-law exponents ($b\neq1$) have been reported in many phenomena including economics \cite{Pareto,Stanley,Stanley2} and family names \cite{family}.
In this section, we extend our model in order to reproduce power-law distributions with other exponents.

The extension is basically modification of the asymmetry of the random walk.
We introduce two types of extension.
In both extensions, the initial and the boundary conditions are the same as in the model in Sec.~\ref{model}.
Extension $A$ of the model evolves as follows:
\begin{enumerate}
\renewcommand{\theenumi}{\roman{enumi}}
\item Choose an entity $i$ randomly from $1\leq i\leq N$. 
\item Add the amount $\alpha_{+} x_i$ or subtract the amount $\alpha_{-} x_i$ randomly  from the chosen entity with the probability $1/2$.
\begin{eqnarray}
x_{i}\rightarrow x_{i}\pm \alpha_{\pm} x_{i}.
\end{eqnarray}
\end{enumerate}
Here the parameters $\alpha_{+}$ and $\alpha_{-}$  are made to be different; 
$\alpha_{+}$ is the growth parameter and $\alpha_{-}$ is the anti-growth parameter.
We restrict ourselves to the case $0<\alpha_{+}<\alpha_{-}<1$;
otherwise, we can have a meaningless solution $b<0$ (See Eq.~(\ref{fbr}) below).

Figure \ref{r0.7} shows some of the results of the simulation of the above procedures for various values of $\alpha$ and $N=10^5$.
\begin{figure}
\begin{center}
\includegraphics[width=0.7\textwidth]{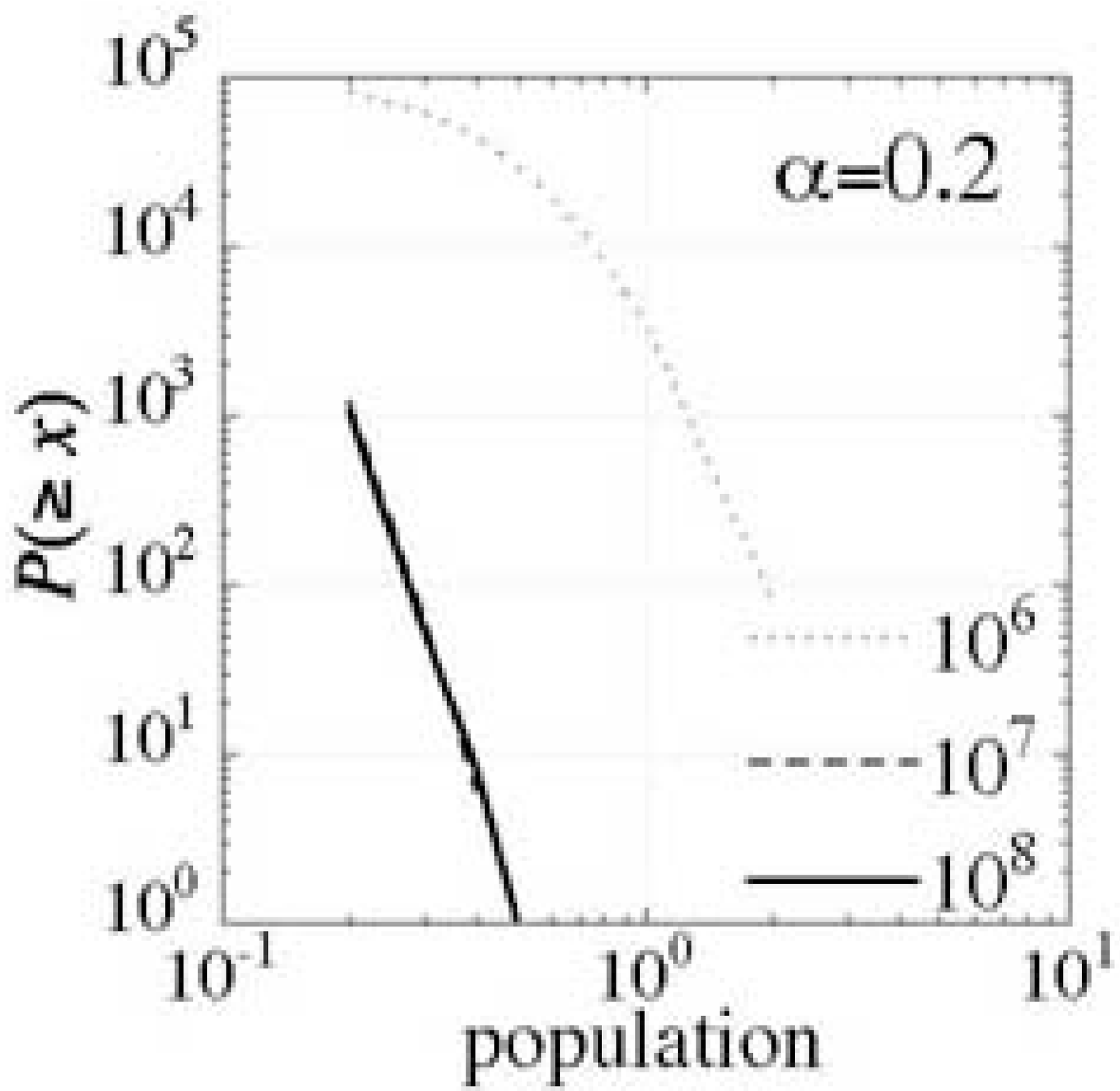}
\includegraphics[width=0.7\textwidth]{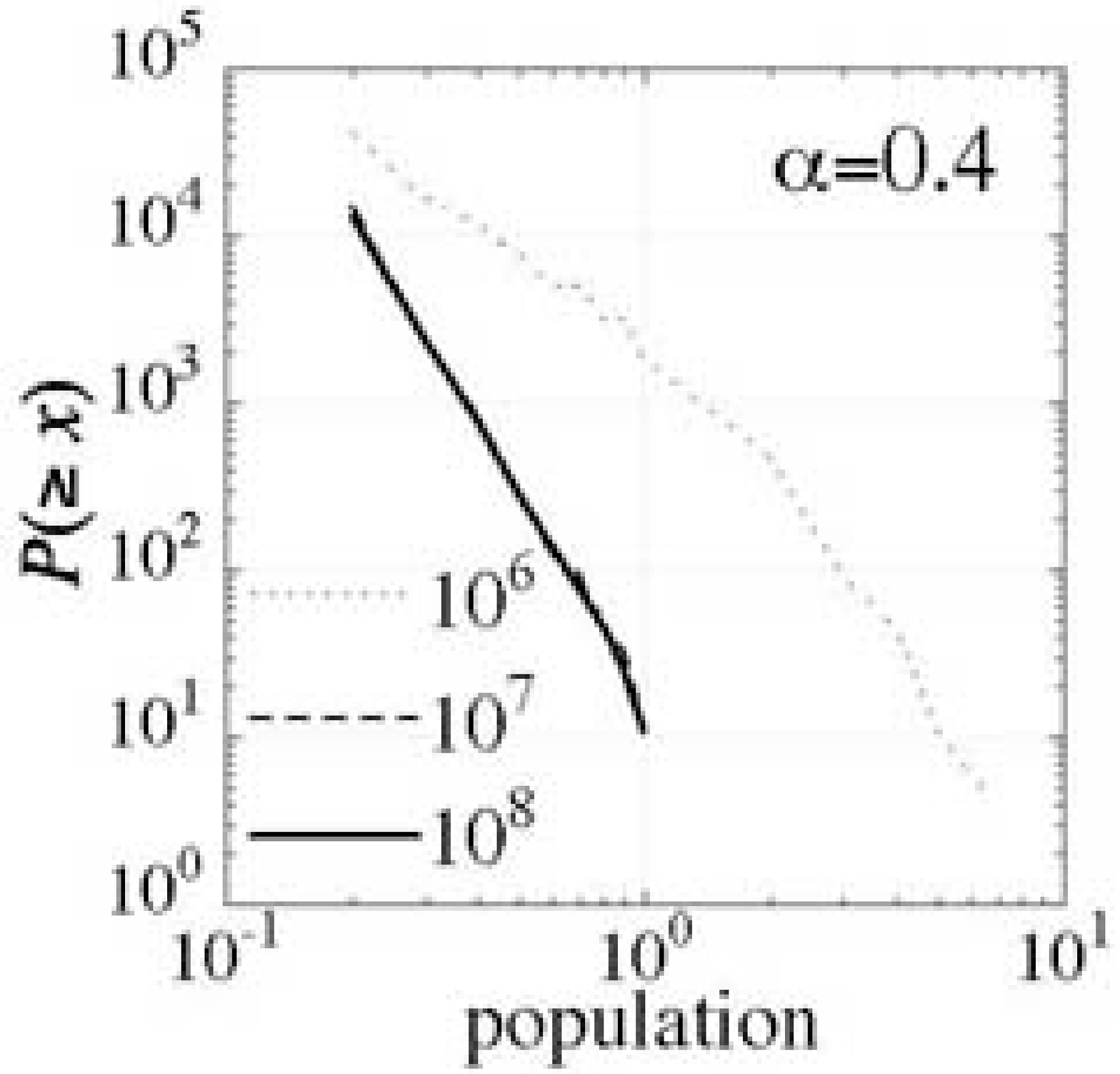}
\end{center}
\end{figure}%
\begin{figure}
\begin{center}
\includegraphics[width=0.7\textwidth]{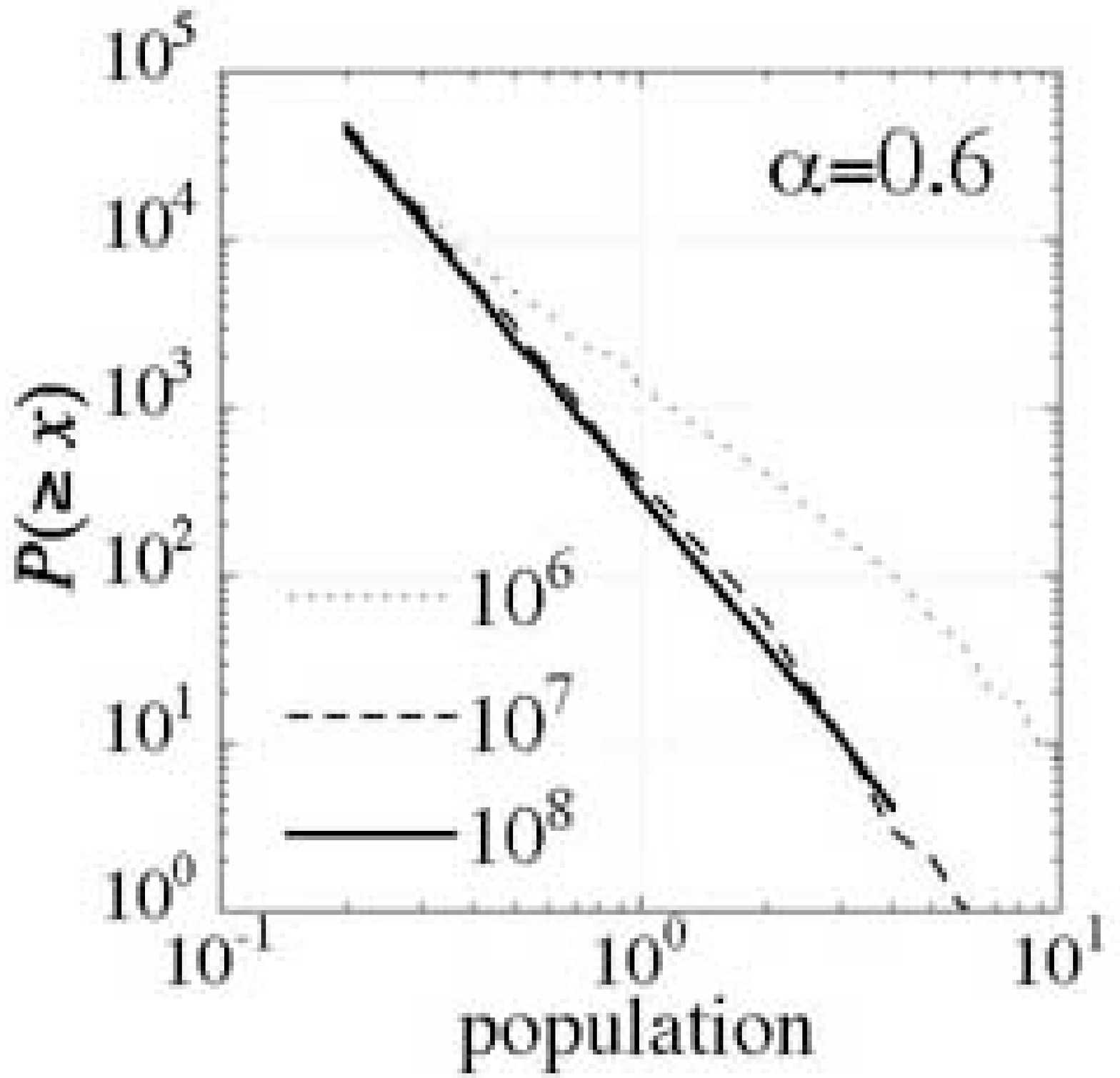}
\includegraphics[width=0.7\textwidth]{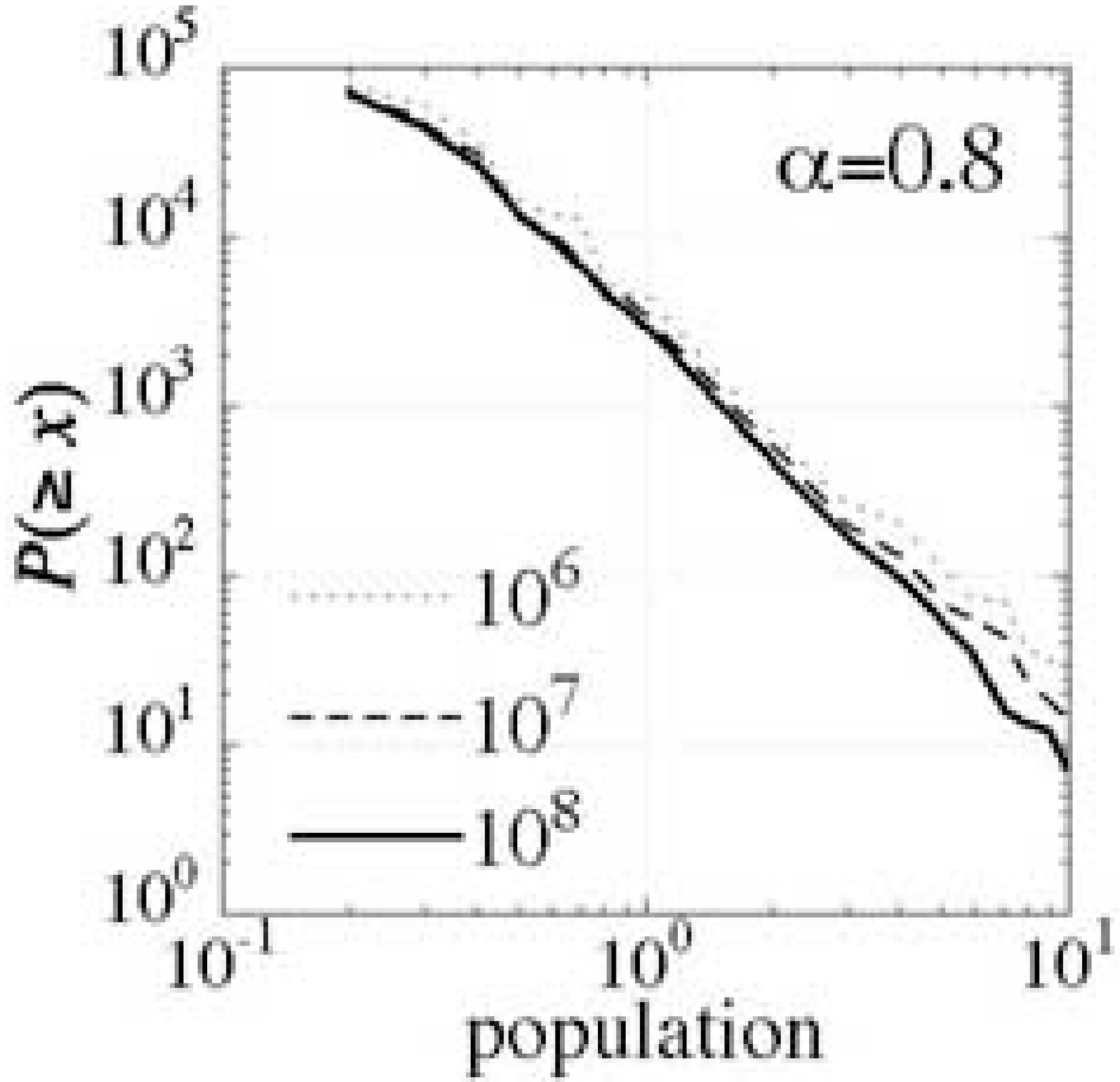}
\caption{Results of the simulation of Extension $A$ for various values of $\alpha$.
 The number of steps is $T=$ $10^6$ (dotted line), $10^7$ (dashed line) and $10^8$ (solid line). 
}
\label{r0.7}
\end{center}
\end{figure}%
We can still see the power-law distribution, but the power-law exponents $b$ changed from unity.
The power-law exponent $b$ takes various values in the modified model while it is fixed to $b=1$ for any values of $\alpha$ in Sec.~\ref{model}. 

We can easily modify the argument in Sec.~\ref{uni} so as to analyze the above modified model, Extension $A$.
Instead of Eq.~(\ref{x}) we now write the time-evolution equation of $p(x)$ as
\begin{equation}\label{r}
\frac{\partial}{\partial t}p(x,t)=-\gamma p(x,t)
+\frac{1}{2}\frac{\gamma}{1-\alpha_{-}}p\left(\frac{x}{1-\alpha_{-}},t\right)
+\frac{1}{2}\frac{\gamma}{1+\alpha_{+}}p\left(\frac{x}{1+\alpha_{+}}, t\right).
\end{equation}
Note that the probability is still conserved.
By rewriting $p$ in terms of $\tilde p$, we transform the evolution equation (\ref{r}) to
\begin{eqnarray}
\frac{\partial}{\partial t}\tilde{p}(\xi ,t)=-\gamma \tilde{p}(\xi,t)
+\frac{\gamma}{2}\tilde{p}(\xi +\beta_{-},t)-\frac{\gamma}{2}\tilde{p}(\xi -\beta_{+},t),
\end{eqnarray}
where 
\begin{eqnarray}\label{betaA}
\beta_{\pm} \equiv \pm\log(1\pm \alpha_{\pm})>0.
\end{eqnarray}
Note the difference between Eq.~(\ref{beta}) and Eq.~(\ref{betaA}).
The equation for the exponent $b$ now reads
\begin{eqnarray}\label{fbr}
f(b)\equiv \frac{1}{2}(1+\alpha_{+} )^b+\frac{1}{2}(1-\alpha_{-})^b=1.
\end{eqnarray}

\begin{figure}
\begin{center}
\includegraphics[width=0.7\textwidth]{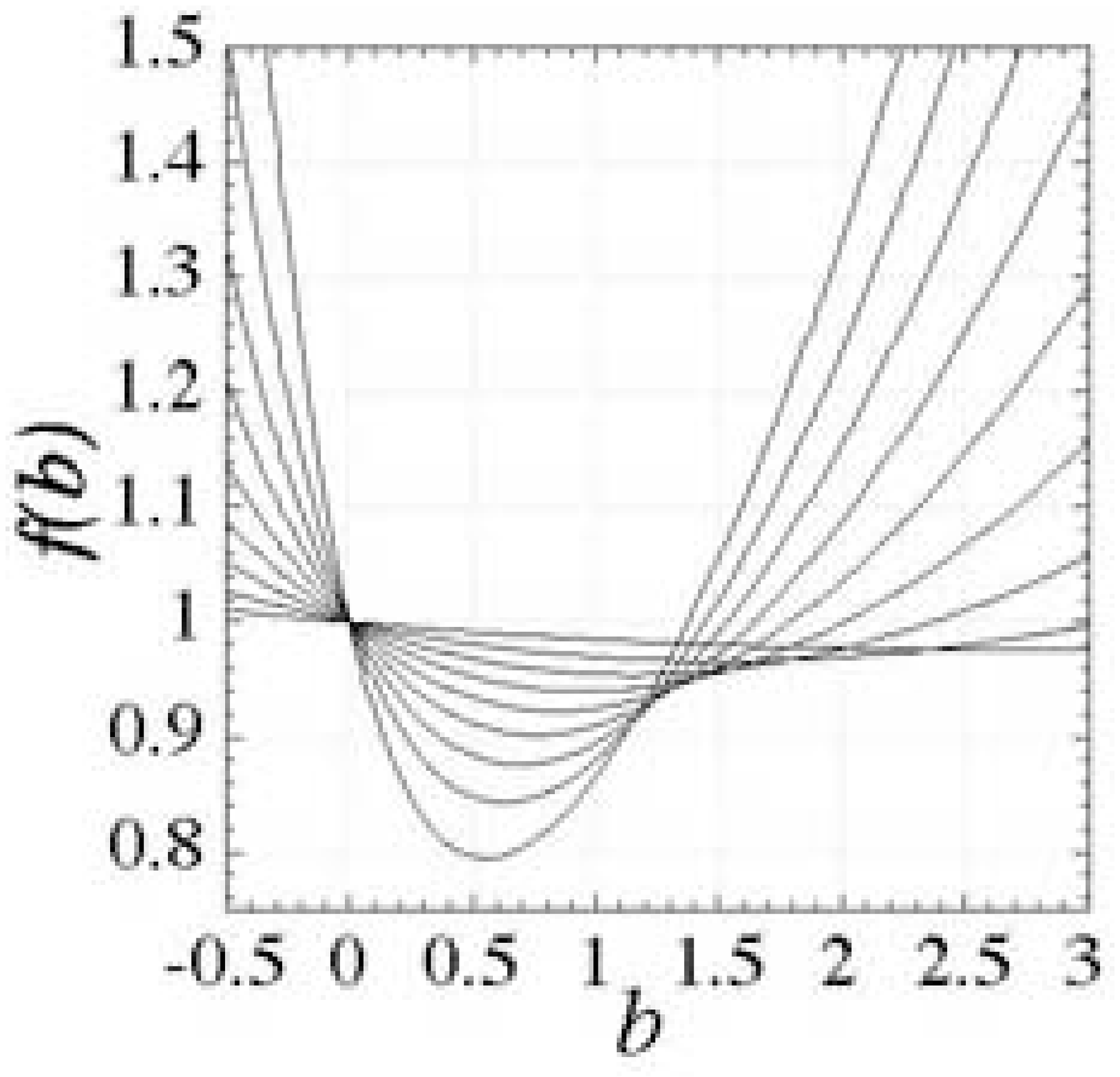}
\caption{
The function $f(b)$ defined in Eq.~(\protect\ref{fbr}) $b$ for various values of $\alpha_{\pm}$ with the ratio $\alpha_{+}/\alpha_{-}=0.7$ fixed.
The curvature of $f(b)$ increases as $\alpha_{\pm}$ become greater.
}
\label{fbr0.7}
\end{center}
\end{figure}

Figure~\ref{fbr0.7} shows the function $f(b)$ for various values of $\alpha_{\pm}$ with $\alpha_{+}/\alpha_{-}=0.7$.
Though the solution $b=0$ is still stable, the other solution depends on $\alpha_{\pm}$. 
Owing to the asymmetry between $\alpha_{+}$ and $\alpha_{-}$, the relevant solution of Eq.~(\ref{fbr}) now depends on $\alpha_{+}$ and $\alpha_{-}$.
\begin{figure}
\begin{center}
\includegraphics[width=0.7\textwidth]{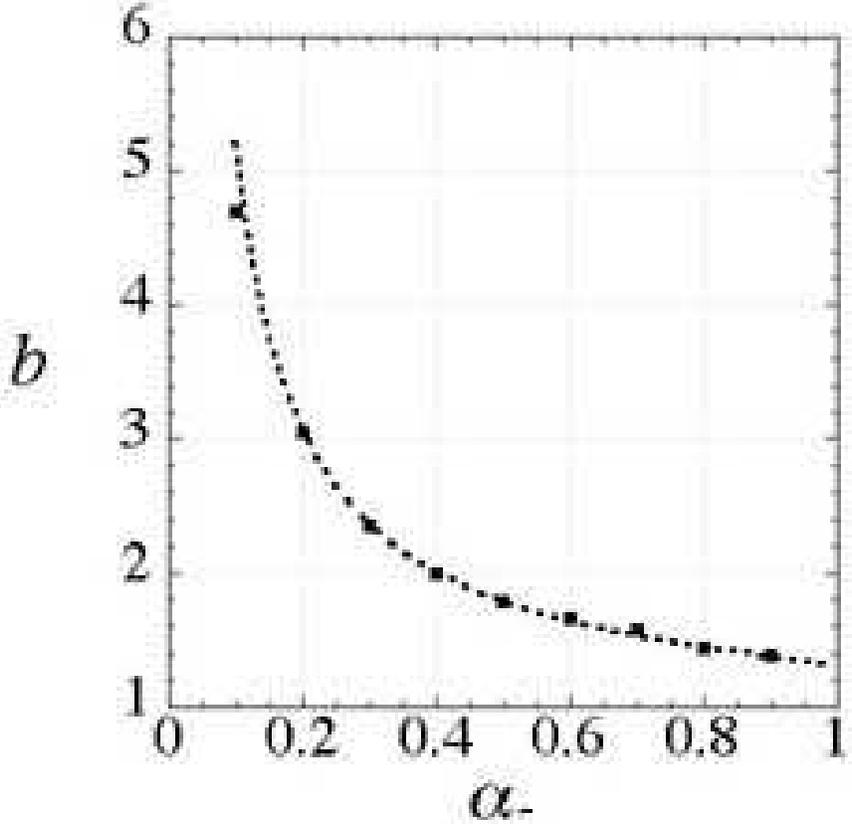}
\caption{The power-law exponent $b$ against the parameter $\alpha_-$ while fixing $\alpha_+/\alpha_-=0.7$.
The solid line shows the solution of Eq.~(\ref{fbr}), while the squares are the simulation results.
}
\label{r0.7alphab}
\end{center}
\end{figure}%
Figure~\ref{r0.7alphab} shows that the simulation results of Extension $A$ (Fig.~\ref{r0.7}) are in good agreement with the solution of Eq.~(\ref{fbr}).

\begin{figure}
\begin{center}
\includegraphics[width=0.7\textwidth]{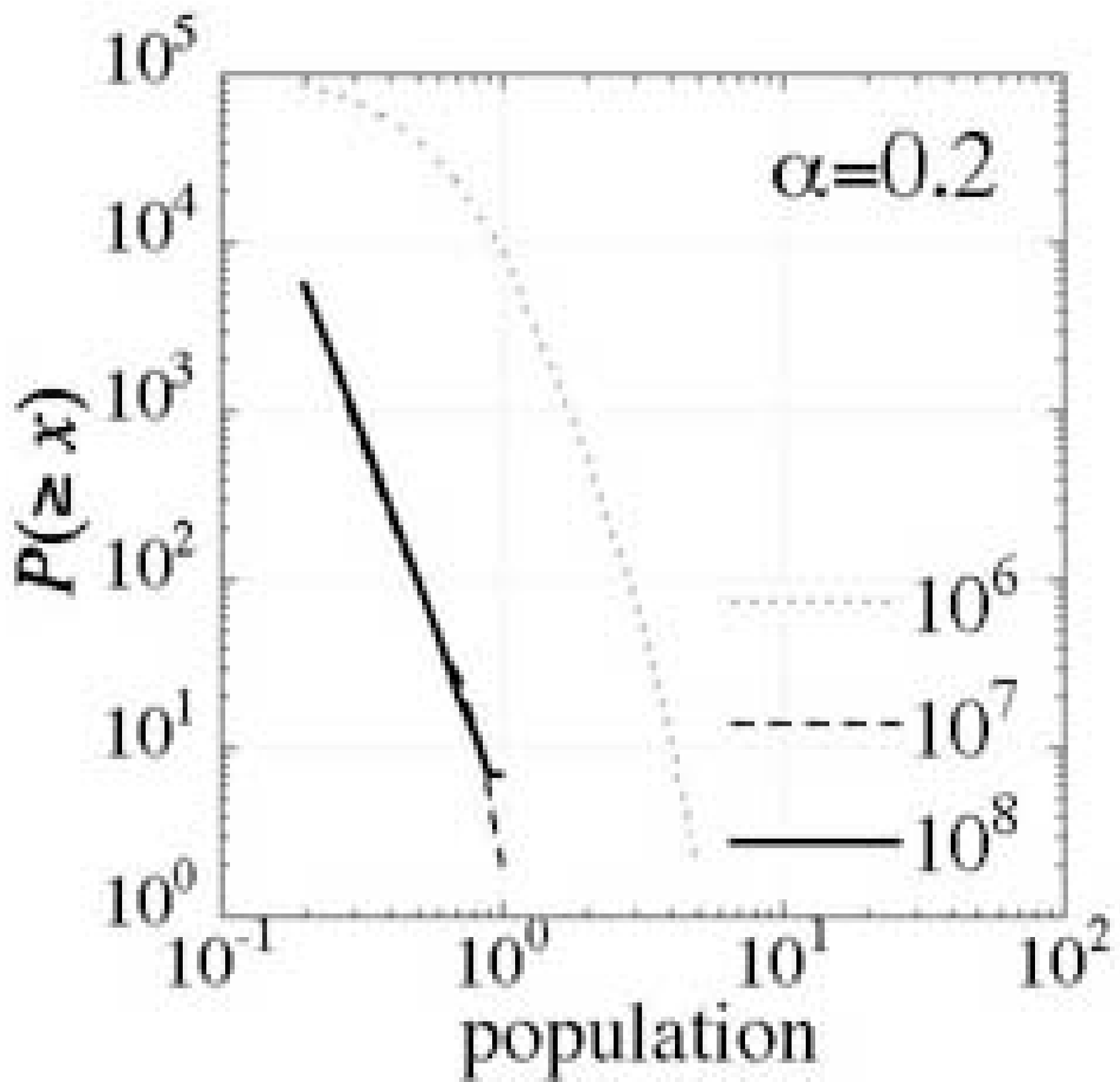}
\includegraphics[width=0.7\textwidth]{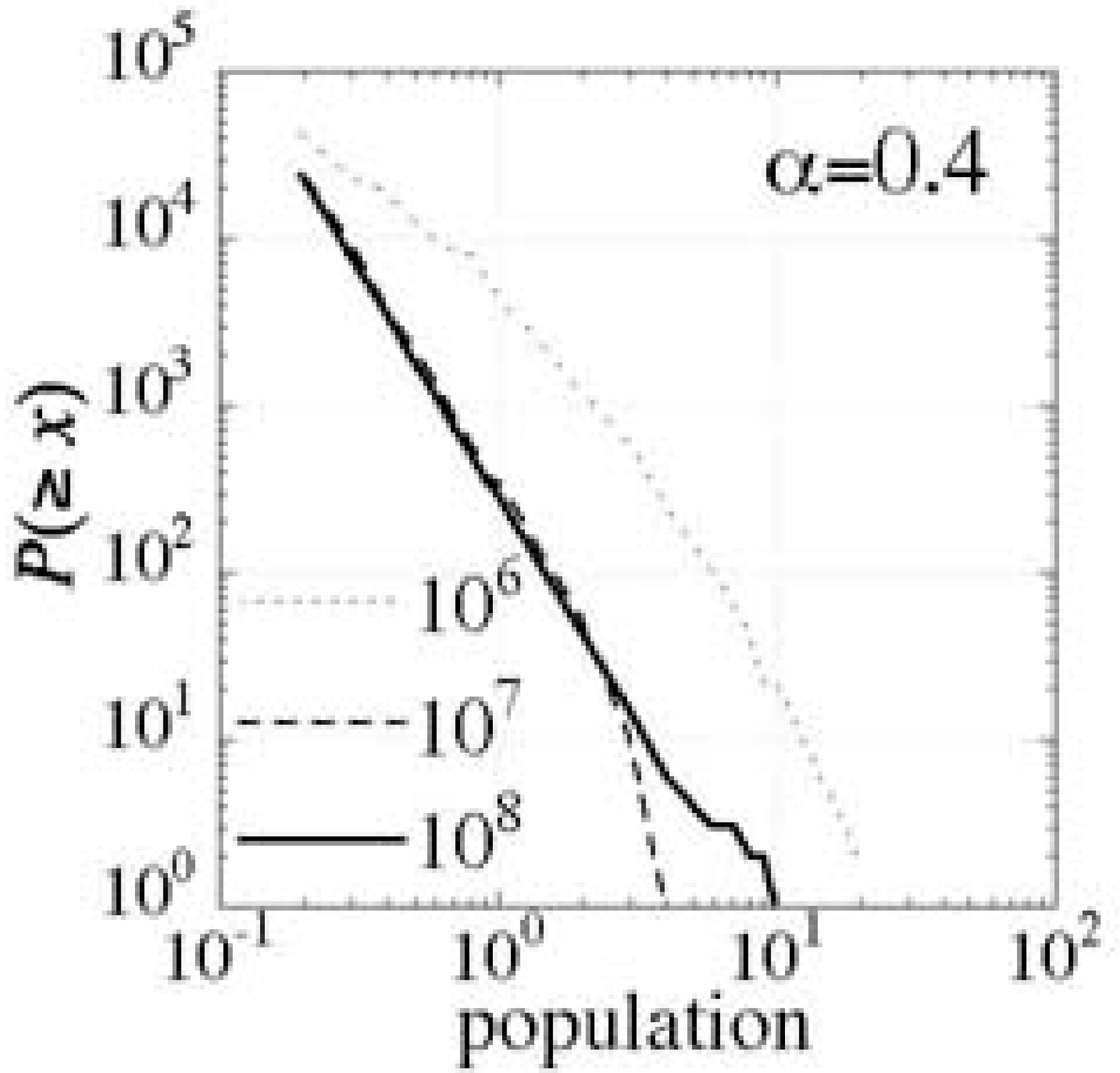}
\end{center}
\end{figure}%
\begin{figure}
\begin{center}
\includegraphics[width=0.7\textwidth]{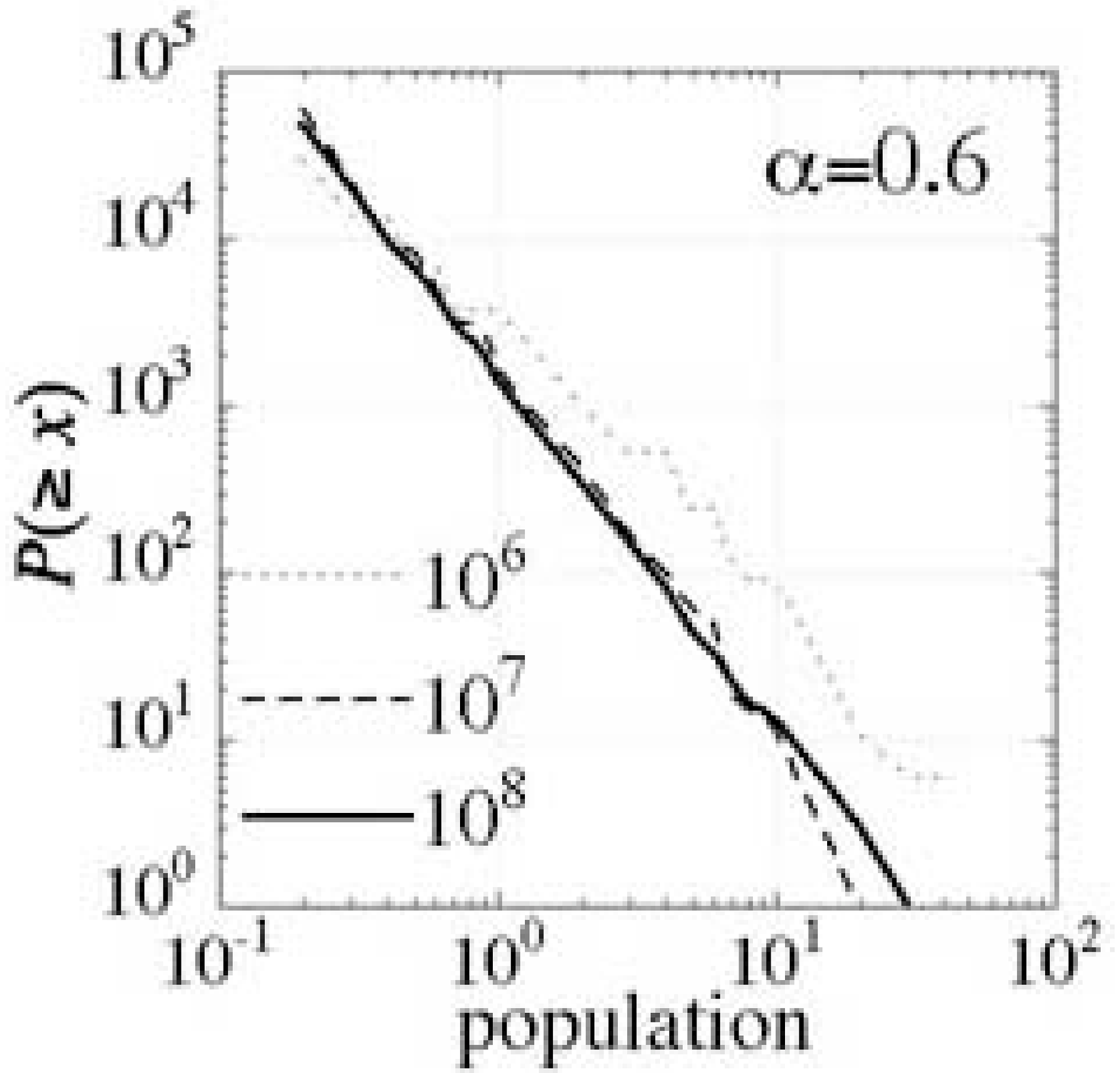}
\includegraphics[width=0.7\textwidth]{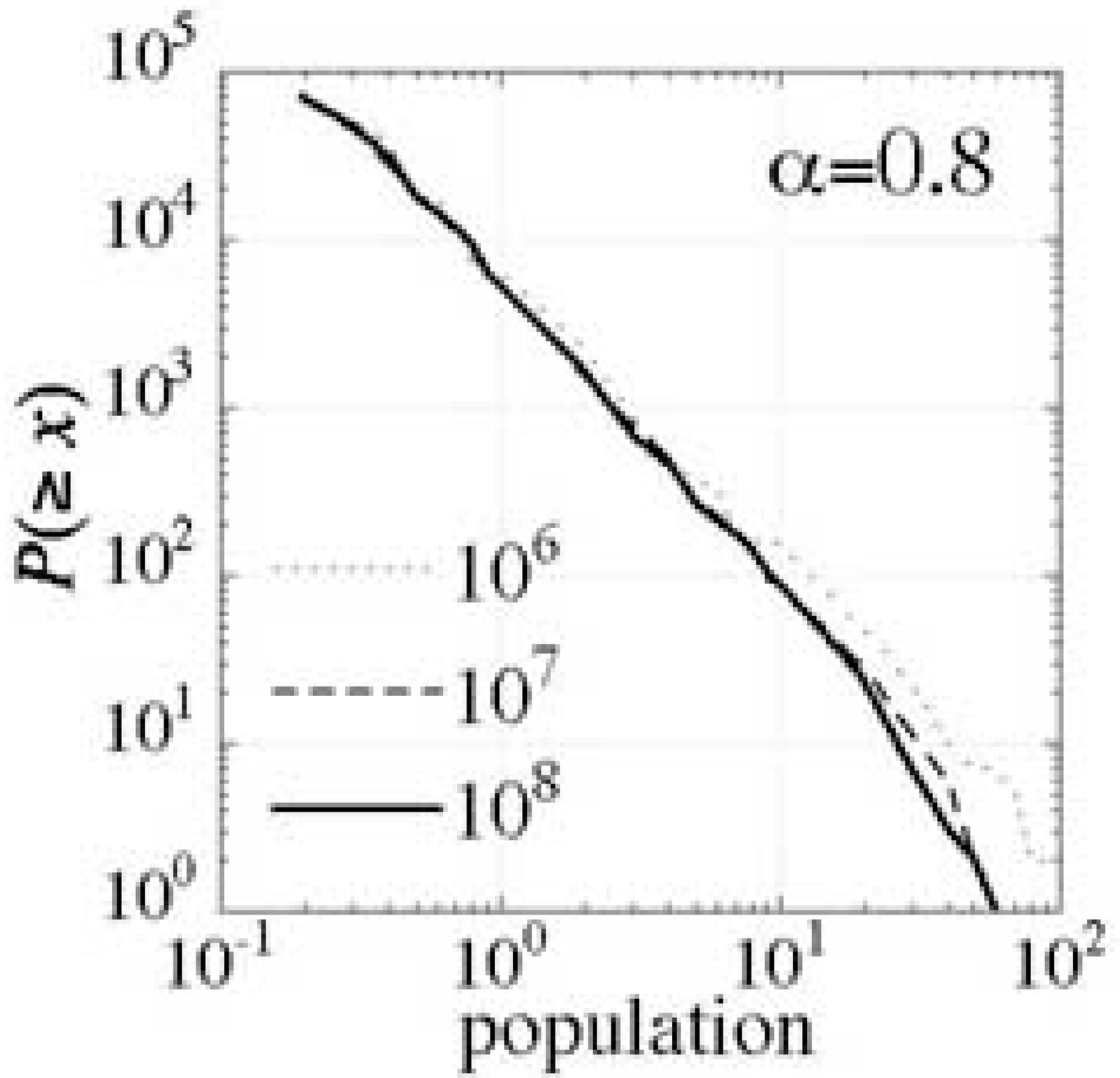}
\caption{Results of the simulation of Extension $B$ for various values of $\alpha$ with $\rho_+ =1/3$ and $\rho_- =2/3$.
The number of steps is $T=$ $10^6$ (dotted line), $10^7$ (dashed line), $10^8$(solid line). 
}
\label{rho1-2}
\end{center}
\end{figure}%
Now we introduce Extension $B$ of the model that also reproduces power-law exponents $b\neq 1$.
Extension $B$  of the model evolves as follows:
\begin{enumerate}
\renewcommand{\theenumi}{\roman{enumi}}
\item Choose an entity $i$ randomly from $1\leq i\leq N$. 
\item Add or subtract randomly the amount $\alpha x_i$ from the chosen entity with the probabilities $\rho_{+}$ and $\rho_{-}$, respectively ($\rho_{-}+\rho_{-}=1$).
In other words,
\begin{equation}
x_{i}\rightarrow 
                      \begin{cases}
                          x_{i}+\alpha x_{i} &  \text{with probability $\rho_+$.}\\
                          x_{i}-\alpha x_{i}  & \text{with prabability $\rho_-$.}
                               \end{cases}
\end{equation}
\end{enumerate}
Figure \ref{rho1-2} shows some of the results of the simulation of Extension $B$.
We again see power laws with $b\neq1$.
The time-evolution equation of $p(x)$ now reads
\begin{eqnarray}\label{rho}
\frac{\partial}{\partial t}p(x,t)=-\gamma p(x,t)+\rho_{-}\frac{\gamma}{1-\alpha}p(\frac{x}{1-\alpha},t)+\rho_{+}\frac{\gamma}{1+\alpha}p(\frac{x}{1+\alpha},t).
\end{eqnarray}
By rewriting $p$ in terms of $\tilde p$, we transform the evolution equation (\ref{rho}) to
\begin{eqnarray}\label{rhoxi}
\frac{\partial}{\partial t}\tilde{p}(\xi,t)&=&-\gamma\tilde p(\xi,t)+\rho_{-}\gamma\tilde p(\xi+\beta_{-},t)+\rho_{+}\gamma\tilde p(\xi-\beta_{+},t),  
\end{eqnarray}
where 
\begin{eqnarray}
\beta_{\pm} \equiv \pm \log(1\pm \alpha)>0.
\end{eqnarray}
The equation for the exponent $b$ now reads
\begin{eqnarray}\label{fbrho}
f(b)=\rho _{-}(1-\alpha)^{b}+\rho_{+}(1+\alpha)^{b}=1.
\end{eqnarray}

\begin{figure}
\begin{center}
\includegraphics[width=0.7\textwidth]{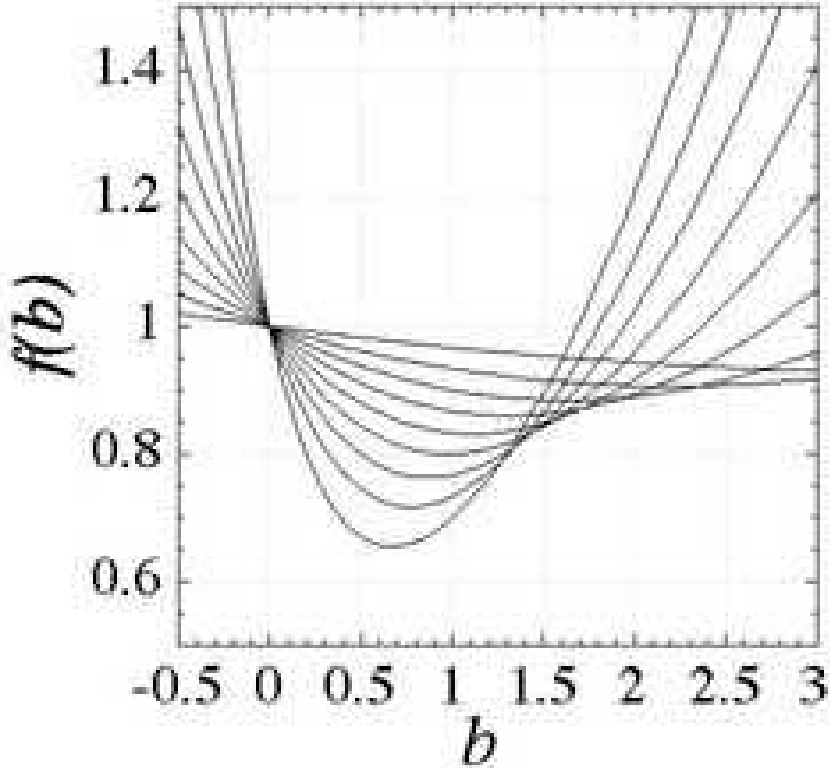}
\caption{
The function $f(b)$ defined in Eq.~(\ref{fbrho}) for various values of $\alpha$ with $\rho_+ =1/3$ and $\rho_-=2/3$ fixed.
The curvature of $f(b)$ increase as $\alpha$ becomes greater.
}
\label{fbrho1-2}
\end{center}
\end{figure}

\begin{figure}
\begin{center}
\includegraphics[width=0.7\textwidth]{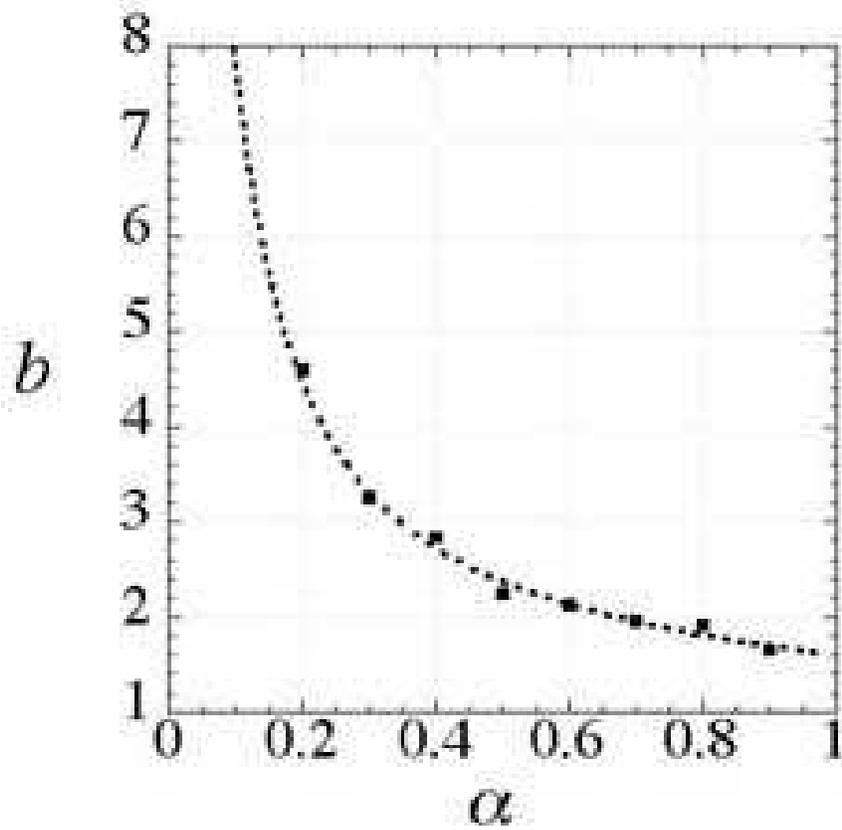}
\caption{The power-law exponent $b$ against the parameter $\alpha$ while fixing $\rho_+/\rho_-=1/2$.
The solid line shows the solution of Eq.~(\ref{fbrho}), while the squares are the simulation results.
}
\label{s12alphab}
\end{center}
\end{figure}

Figure~\ref{fbrho1-2} shows the function $f(b)$ for various values of $\alpha$ with $\rho_{+}/\rho_{-}=1/2$.
Though the solution $b=0$ is still stable, the other solution depends on $\alpha$. 
Owing to the asymmetry between $\rho_+$ and $\rho_-$, the relevant solution of Eq.~(\ref{fbrho}) depends on $\alpha$, $\rho_+$ and $\rho_-$.
Figure~\ref{s12alphab} shows that the simulation results of Extension $B$ (Fig.~\ref{rho1-2}) are in good agreement with the solution of Eq.~(\ref{fbrho}).
\section{Summary}
In the present paper, we explained the origin of the universality of Zipf's law theoretically and demonstrated it numerically.
We claim that the essence of Zipf's law is a size change of each entity proportional to its size.
The simulation results of our simple and generic model suggest that the present explanation of Zipf's law is applicable to various phenomena both in natural and social sciences.

We also extended the model slightly and reproduced power laws other than Zipf's law.
We revealed that some asymmetry between the increase and the decrease causes the deviation from Zipf's law.
Equations (\ref{fbr}) and (\ref{fbrho}) may be useful in estimating parameters in actual power-law phenomena.

\appendix
\section{Exponential Solutions}
In this appendix, we discuss the validity of the assumption (\ref{sp}) for the solution of Eq.~(\ref{ss}).
Carrying out the Fourier transform of Eq.~(\ref{ss}), we obtain 
\begin{equation}\label{f-ps}
-2 \widetilde{\widetilde {p_{s}}}(k)
+e^{ik\beta_{+}}\widetilde{\widetilde {p_{s}}}(k)
+e^{-ik\beta_{-}}\widetilde{\widetilde {p_{s}}}(k)=0,
\end{equation}
or
\begin{equation}\label{eik}
e^{ik\beta_{+}}+e^{-ik\beta_{-}}=2,
\end{equation}
where 
\begin{equation}\label{fourier}
\widetilde{\widetilde {p_{s}}}(k)=\int \tilde p_s(\xi)e^{ik\xi}d\xi.
\end{equation}
By substituting Eq.~(\ref{beta}), we can write Eq.~(\ref{eik}) in the form
\begin{equation}\label{ik}
(1+\alpha)^{ik}+(1-\alpha)^{ik}=2.
\end{equation}

Let us look for solutions $k$ in the complex plane by writing
\begin{equation}\label{ReIm}
k=k_r-ib,
\end{equation}
where $k_r$ and $b$ are real numbers.
If we have solutions with $k_r=0$ only, the assumption Eq.~(\ref{sp}) is justified.
Substituting Eq.~(\ref{ReIm}) for $k$ in Eq.~(\ref{ik}), we have
\begin{equation}\label{ekb}
(1+\alpha)^{b}e^{ik_{r}\beta_{+}}+(1-\alpha)^{b}e^{-ik_{r}\beta_{-}}=2.
\end{equation}
To satisfy Eq.~(\ref{ekb}), we need the phase factors to be real:
\begin{equation}\label{ee1}
e^{ik_r\beta_+}=e^{-ik_r\beta_-}=1
\end{equation}
or
\begin{equation}\label{e-1e1}
-e^{ik_r\beta_+}=e^{-ik_r\beta_-}=1.
\end{equation}
In the case of Eq.~(\ref{ee1}), we have
\begin{equation}\label{krmn}
k_r=\frac{2m\pi }{\beta_+}=\frac{2n\pi }{\beta_-},
\end{equation}
while in the case of Eq.~(\ref{e-1e1}) we have 
\begin{equation}\label{krm-1n}
k_r=\frac{(2m+1)\pi}{\beta_+}=\frac{2n\pi}{\beta_-},
\end{equation}
where $m$ and $n$ are integers.
In order for the solutions (\ref{krmn}) and (\ref{krm-1n}) to exist, the ratio of the step sizes $\beta_+$ and $\beta_-$ must be a rational number.
It means that the random walker has a possibility of returning exactly to the starting point.
We consider that such cases are quite exceptional;
if we choose a general value of $\alpha$, the ratio $\beta_+/\beta_-$ is an irrational number in general.
Hence we neglect the solutions (\ref{krmn}) and (\ref{krm-1n}) and assume $k_r=0$.
Therefore the remaining solutions have the form $k=-ib$.
We can thus justify the assumption (\ref{sp}).


\begin{thebibliography}{9}
\bibitem{Zipf}
 G. K. Zipf: \textit{Human Behavior and the Principle of Least Effort} (Addison-Wesley, Cambridge, 1949).
\bibitem{population}
For example, M. Marsili and Y.-C. Zhang: Phys. Rev. Lett. \rm{\bf80} (1998) 2741.
\bibitem{Takayasu}
K. Okuyama, M. Takayasu and H. Takayasu: Physica A \rm{\bf269} (1999) 125.
\bibitem{Takayasu2}
   H. Takayasu, M. Takayasu, M. P. Okazaki, K. Marumo and T. Shimizu: in 
  \textit{Paradigms of Complexity}, ed. M. M. Novak (World Scientific, Singapore 2000)  p. 243.
 \bibitem{Render}
 F. Leyvraz and S. Render: Phys. Rev. Lett. \rm{\bf 88} (2002) 068301
 \bibitem{Sornette}
 D. Sornette: Phys. Rev. E \rm{\bf 57} (1998) 4811
 \bibitem{kenji}
 K. Kawamura and N. Hatano: J. Phys. Soc. Jpn. \rm{\bf71} (2002) 1211.
 \bibitem{internet}
 S. Lawrence and C.L. Giles: Science \rm{\bf280} (1998) 98.
 \bibitem{traffic}
 D.L. Gerlough and M.J. Huber: \textit {Traffic flow theory} (National Research Council, Wasington, DC, 1975).
 \bibitem{Pareto}
 V. Pareto: Le Cours d\lq Economie Politique. (Macmillan, London, 1897.)
 \bibitem{Stanley}
 M.H.R. Stanley, L.A.N. Amaral, S.V. Buldyrev, S. Havlin, L. Leschhron, P. Maass, M.A. Salinger and H.E. Stanley: Nature \rm{\bf 379} (1996) 804.
\bibitem{Stanley2}
 Y. Lee, L.A.N. Amaral, D. Canning, M. Meyer, and H.E. Stanley: Phys. Rev. Lett. \rm{\bf81} (1998) 3275.
 \bibitem{fish}
H. S. Niwa: J. Theor. Biol. \rm{\bf 195} (1998) 351.
 \bibitem{family}
 S. Miyazima, Y. Lee, T. Nagamine and H. Miyajima: Physica A \rm{\bf 278} (2000) 282.
\end{thebibliography}
\end{document}